\documentclass[preprint,12pt]{elsarticle}

\usepackage{amsmath}
\usepackage{amssymb}
\usepackage{graphicx}% Include figure files
\usepackage{dcolumn}
\usepackage[bottom]{footmisc}% Align table columns on decimal point
\usepackage{bm}
\usepackage{bbm}
\usepackage{mathtools}
\usepackage{breqn}
\usepackage{amsfonts}
\usepackage{enumitem}
\usepackage{float}
\usepackage{graphicx}
\usepackage{slashed}
\usepackage{breqn}
\usepackage{float}
\usepackage{pifont}
\usepackage{multirow}

\usepackage{bm}
\usepackage{physics}
\usepackage{listings}
\usepackage{algorithm2e}
\usepackage{xcolor,mdframed}
\journal{Elsevier}

\begin{document}

\begin{frontmatter}



\title{Lattice Fermionic Casimir effect in a Slab Bag and Universality}


\author{Yash V. Mandlecha \footnote{\textit{Email address:} \texttt{yash17@iiserb.ac.in}, \texttt{mandlec1@msu.edu} 
%Address after 16 August 2022 : Department of Physics and Astronomy, 567 Wilson Rd, East Lansing, MI 48824, United States
}}
%\email{yash17@iiserb.ac.in} 
%\address{Department of Physics,
%   Indian Institute of Science Education and Research Bhopal, 
%    Bhopal 462066, Madhya Pradesh, India}

\author{Rajiv V. Gavai \footnote{\textit{Email address:}  \texttt{gavai@tifr.res.in, rajiv@iiserb.ac.in}}}
\address{Department of Physics,
    Indian Institute of Science Education and Research Bhopal, 
    Bhopal 462066, Madhya Pradesh, India}

%\email{rajiv@iiserb.ac.in} 
\begin{abstract}
%The Casimir effect for photons and Dirac fermion fields was studied and generalized to ($D+1$)-dimensional spacetime in the continuum.
%  It is observed from existing literature that Casimir energy for the Wilson and overlap fermions exactly matches the continuum expressions for the Dirac fermionic Casimir effect in the zero limit of the lattice spacing. However,
 %The  model is applied to treat the slab with perfectly conducting parallel plates as a bag with confined fermion fields.
 We apply the physically more appealing MIT Bag boundary conditions to study the Casimir effect 
 on the lattice. 
 Employing the formalism of  Ref. \cite{Ishikawa:2020ezm} to calculate the Casimir energy for free lattice fermions, we show that the results for 
  the naive, Wilson and overlap fermions 
  match the continuum expressions precisely in the zero lattice spacing limit, as expected from universality.
  In contrast to Ref. \cite{Ishikawa:2020ezm} where the result for the naive fermions rapidly oscillates with the lattice size for both, the periodic (P) and anti-periodic (AP) boundary conditions, no oscillations are observed with the lattice size. 
  Furthermore, the apparent violation of the universality for naive fermion in Ref. \cite{Ishikawa:2020ezm} is shown to be cured 
  by applying suitable series extrapolation techniques, thus demonstrating     
  that the Casimir energy for the naive fermions with periodic/antiperiodic boundary conditions agrees with the results for
  other free lattice fermions, and can be used to obtain the results for the Dirac fermion in the zero limit of the lattice spacing.
\end{abstract}



\begin{keyword}
Casimir effect \sep lattice fermions \sep MIT Bag model
\end{keyword}

\end{frontmatter}

\newpage
\section{\label{sec:level1} Introduction}

 %The Quantum vacuum is filled with continuously fluctuating fields. The virtual particles can be created in and annihilated back to the vacuum for a time interval dictated by the uncertainty relation:
%\begin{equation}
%\Delta E \cdot \Delta t \geq \frac{\hbar}{2}
%\end{equation}
% The total zero-point energy of the vacuum, in which the spectrum of possible wave modes forms a continuum, is a divergent quantity obtained from the standard approach of canonical field quantization. 
  The Casimir effect \cite{Casimir:1948dh} arises as a consequence of restricting the allowed  
  %presence of macroscopic bodies like the parallel plates introduces non-trivial boundary conditions into the system.
 %This modifies
  zero-point fluctuations in the vacuum to discrete values due to the experimental set-up.
 %and restricts the normal component of the allowed wave modes between the plates to discrete values.
 This vacuum distortion manifests as an attractive force
 %between the parallel plates 
 due to a finite lowering of the vacuum energy. These forces have been measured between two parallel plates \cite{Bressi}, and the phenomenon was confirmed to a high degree of precision in several experiments \cite{Lamoreaux, Mohideen} recently. 
 One might conclude quantum phenomena like these to be esoteric, with a limited practical consequence. But, as characteristic distances get smaller, their effects have become 
 %very relevant and
 increasingly significant in nanotechnology, for example, the silicon integrated
circuits based on micro-and nano-electromechanical systems \cite{Gong}. The Casimir effect
also established the fact that energy density in certain regions of space is
negative relative to the ordinary vacuum energy. This has exciting consequences,
as such effects contribute to the stability of hadrons and model
colour-confinement \cite{Johnson:1975zp, Milton:2001yy, ChernodubNonPert}, might
make it possible to stabilize a traversable wormhole \cite{Morris}, and provide
significant insights into the cosmological constant problem \cite{MAHAJAN20066}.
The  Casimir effect has also been generalized to the scalar and fermion
fields.

 The strong coupling between quark and gluon fields gives rise to various
non-perturbative phenomena in Quantum Chromodynamics. Given the role Casimir
effect plays in the MIT Bag model, it would be interesting to investigate its
role in the confinement of quarks in hadrons and spontaneous symmetry breaking
using the formalism defined in Ref. \cite{Ishikawa:2020ezm}. Clearly using a
spacetime lattice for such investigations appears to be a natural choice.  An
important motivation for studying the Casimir effect for various lattice
fermions is that their form also appears in condensed matter systems, like Dirac
semi-metals, topological insulators and ultracold atom systems, which are
currently active areas of research. Chern Insulators are also shown to exhibit a repulsive Casimir effect
 %. If this problem is solved in generality, it can help us understand its
 %consequence to the various condensed matter systems in a better fashion
  \cite{Araki:2013qva, Rodriguez}. The negative mass Wilson
fermions and overlap fermions
  %with \textcolor{red}{MDW} kernel
  correspond to the bulk and surface fermions of the topological insulator,
respectively \cite{Ishikawa:2020icy}. 
  %They also exhibit oscillation of Casimir energy on odd and even lattice sizes
  %in certain phases \cite{Ishikawa:2020icy}.
  Studying the Casimir effect for these lattice fermions \cite{Susskind} is equivalent to
studying it for corresponding topological insulators, which can then be experimentally observed for very small lattice sizes.

The Casimir effect for free lattice fermions was first studied in Ref. \cite{Ishikawa:2020ezm, Ishikawa:2020icy} using periodic and antiperiodic boundary conditions. Periodic and anti-periodic boundary conditions are are often used as a standard theoretical setup in 
lattice simulations or condensed matter physics. While
the Casimir energy for the Wilson and overlap fermions exactly matched the continuum expressions expected for the fermionic Casimir effect in the zero limit of the lattice spacing, it was observed that the result for the naive fermion rapidly oscillated and approached two different expressions in the continuum limit for the odd and even lattice sizes respectively. This lead Ref. \cite{Ishikawa:2020ezm} to claim  that the naive fermion cannot be used to compute Casimir energy for the Dirac fermion, which is an apparent violation of the universality,

Motivated by the physically appealing fact that the MIT Bag boundary conditions
\cite{Johnson:1975zp} prevent the fermionic current from crossing the plates,
thereby ensuring the confinement of fermions within the bag, we adopt these
boundary conditions in this work to study the fermionic Casimir energy. If the
fermion field $\psi$ is subjected to the MIT Bag boundary conditions, and
confined between the two parallel plates placed at $x^1 = 0$ and $x^1 = d$, 
$\psi$ satisfies the following equation
\begin{equation}
\label{bound}
(1+i\gamma^1)\psi|_{x^1 =0,d}= 0~.
\end{equation}
%The conditions for the parallel plates placed at $x^1=0$ and $x^1=d$ are respectively:
%\begin{equation}
%(1 \mp i \gamma^1)\psi = 0
%\end{equation}
%using the chiral representation of the Dirac matrices, where $j = 1,\dots,D$ and $\sigma_j$ are the pauli matrices satisfying $\{\sigma_j,\sigma_l\} = \sigma_j\sigma_l + \sigma_l\sigma_j = 2\delta_{lj}$ .
%\\ The standard positive and negative frequency solutions of the Dirac Equation respectively can be written as :
 %\begin{equation}
%  \label{varphi1}
% \psi^{(+)} = e^{-i\omega t}\left( \begin{array}{cc}
%\varphi^{(+)} \\
%\dfrac{-i\bm{\sigma\cdot\nabla}\varphi^{(+)}}{\omega +m}
%\end{array} \right), \;\;\text{and  } 
%
% \psi^{(-)} = e^{i\omega t}\left( \begin{array}{cc}
%\dfrac{i\bm{\sigma\cdot\nabla}\varphi^{(-)}}{\omega +m}\\
%\varphi^{(-)} 
%\end{array} \right),
%
% \end{equation}
 %where $\bm{\sigma} = (\sigma ^1,...,\sigma ^{D})$ and the spinors $\varphi^{(+)}$ and $\varphi^{(-)}$ correspond to the particle and antiparticle fields respectively, are given by the expression :
 %\begin{equation}
  %   \varphi^{(\pm)} = ( \varphi_+^{(\pm)} e^{ik_1x^1} + \varphi_-^{(\pm)} e^{-ik_1x^1}) \exp{\pm i\sum_{j=2}^D k_jx^j}
  %   \end{equation}
By substituting the standard positive and negative frequency solutions of the
Dirac equation in the equation for the MIT Bag boundary conditions one obtains the
following transcendental relations for the corresponding momentum $k_1$
\cite{Mamaev:1980jn, BelussiSaharian}: \begin{equation} \label{trans}
md~\sin(k_1d)/k_1d + \cos(k_1d) = 0.  \end{equation} For the massless fermion
case, this reduces to \begin{equation} \label{allow} k_1d =
\left(n+\frac{1}{2}\right)\pi  \;\;\;n\in \mathbb{Z} \end{equation} These
boundary conditions will be used to calculate Casimir energy for the lattice
fermions in section (\ref{slabbag}).

The results we obtained from the MIT Bag boundary
% in
%continuum, to be compared with our expressions calculated for free lattice
%fermions \cite{Susskind}. 
conditions were encouraging and precisely matched the continuum results for all
lattice fermions, including the naive fermions, in the zero limit of the lattice
spacing. As expected, doubling was observed for the results for the naive
fermion. In order to investigate whether the choice of boundary conditions
is crucial for our results above, we focused our attention on the periodic and
antiperiodic boundary conditions for naive fermions again. Instead of
examining the even and the odd lattice size series separately we treat them
together as one series and applied known series extrapolation techniques.
%So far, for simplicity this study is done only for free lattice
%fermions. Studying its consequences on the confinement of fermion fields using
%the MIT bag boundary conditions will have more physical meaning once we
%introduce interactions between the fermion and gauge fields. Although, the
%studies of qualitative properties using free lattice fermions is a fundamental
%step and will be useful to study the Casimir effect for interacting lattice
%fermions in future. 
Contrary to the claim of Ref. \cite{Ishikawa:2020ezm} of an apparent violation 
of the universality for the naive fermion, we succeeded in demonstrating
numerically  that the Casimir effect for Dirac fermions can also be computed
from the naive fermion in the zero limit of the lattice spacing since looked upon this way, the series converges to the same result as given by other lattice fermions
in the continuum limit. We mention the known continuum results for both massless 
and massive fermions for the MIT Bag boundary conditions for reference and comparison.

\subsection{Massless fermions in $(3+1)$-dimensional spacetime}
 In continuum theory, one computes the zero-point energy using dimensional regularization by summing over the odd integer modes obtained in (\ref{allow}). %There are two spin modes for fermions. An additional factor of two comes from the particle and antiparticle. Fermions also have a characteristic negative sign. 
 We obtain the characteristic Casimir energy for massless fermions in (1+1)- and  (3+1)-dimensional spacetime as \cite{Paola, Milton:2001yy} ($\hbar=c=\kappa_B$=1):
\begin{equation}
 \label{fercont}
E^{3+1,\text{cont},f}_{\text{Cas}}(d) = -\frac{7\pi^2 }{2880 d^3}\;\;;\;\;E^{1+1,\text{cont},f}_{\text{Cas}}(d) = -\frac{\pi}{24 d}
\end{equation}
%A general expression for the massless fermionic Casimir energy in the $D+1$-dimensions is also obtained.
 
  \subsection{Massive fermions in $(3+1)$-dimensional spacetime}  
 %The Generalized Abel-Plana formula \cite{Saharian:2006iv} is employed to renormalize the expression to obtain a finite result for the Casimir energy. The integral obtained for massive fermionic fields has no closed expression in terms of standard functions.
 One obtains approximate solutions for the Casimir energy of massive fermions (\textit{mf}) in continuum theory for two limiting cases, namely $md \ll 1$ and $md \gg 1$. The Casimir force per unit area in the limit $md \gg 1 $, which we shall verify for the lattice fermions is \cite{ELIZALDE, Cruz}:
% \begin{equation}
 %     {E^{3+1,\text{cont},mf}_{\text{Cas}}(d)} = -\frac{7}{4}\frac{\pi^2\hbar c}{720d^3}  \left(1-\frac{120md}{7\pi^2}\right)
 %\end{equation}
%Now, in the limiting case $md \gg 1$ the Casimir energy is given by :
 \begin{equation}
 \label{massivefer}
     E^{3+1,\text{cont},mf}_{\text{Cas}}(d)  = -\frac{3}{32}\sqrt{\frac{m}{\pi^3d^5}}e^{-2md}
 \end{equation}
 where the Casimir energy decays exponentially with $md$. 
%\section{Formalism for Casimir energy on lattice} 
%    The lattice representation provides a cut-off for the removal of ultraviolet infinities in a Quantum Field Theory.  The conventional regularization schemes are employed to remove the divergences met during the calculation of processes through a Feynman diagram. The lattice regularization serves the purpose of a non-perturbative cut-off which goes beyond the diagrammatic approach. 
    %The lattice regularization directly eliminates all wavelengths less than twice the lattice spacing `$a$', which is also the case when we cut off all frequencies below $k_m$ as discussed in section (\ref{sec:level2}). 
  %  The physics of a particular system can be extracted from the numerical calculations in the continuum limit when the lattice spacing limit tends to zero. \\  
   %The theoretical construction of the lattice fermions relates to the Nielsen-Ninomiya theorem. The `doubler' particles appear as additional degrees of freedom in the presence of hermiticity, chiral symmetry, locality and translational invariance in the naive discretization of space.

   \section{Casimir effect for lattice fermion fields in slab-bag}
\label{slabbag}

%    The lattice representation provides a cut-off for the removal of ultraviolet infinities in a Quantum Field Theory. The conventional regularization schemes are employed to remove the divergences met during the calculation of processes through a Feynman diagram. The lattice regularization serves the purpose of a non-perturbative cut-off which goes beyond the diagrammatic approach. 
    %The lattice regularization directly eliminates all wavelengths less than twice the lattice spacing `$a$', which is also the case when we cut off all frequencies below $k_m$ as discussed in section (\ref{sec:level2}). 
  %  The physics of a particular system can be extracted from the numerical calculations in the continuum limit when the lattice spacing limit tends to zero. \\  
   %The theoretical construction of the lattice fermions relates to the Nielsen-Ninomiya theorem. The `doubler' particles appear as additional degrees of freedom in the presence of hermiticity, chiral symmetry, locality and translational invariance in the naive discretization of space.
   In this letter, out of the $D$ latticized spatial dimensions, we consider only one 
   ($x^1$) compactified by a boundary condition at $x^1=0,\;d$. The corresponding spatial momentum component $p_1$ is discretized, while the other momenta components remain continuous. Initially, modelling time is kept continuous and temporal components of momentum are unaffected by latticization. 
% In space where the spatial direction is compactified, the  spatial momentum component ($p_1$) is discretized.
 The non-zero lattice spacing is `$a$', with an ultraviolet cut-off scale `$1/a$' in momentum space. The energy - momentum dispersion relation for lattice fermions is obtained from the Dirac operator defined in a relativistic fermion action. The energy is defined as:
\begin{equation}
\label{Drel}
a E(ap) = a \sqrt{\mathcal{D}^{\dagger} \mathcal{D}}  
\end{equation}
 The Dirac operator, $\mathcal{D}$ includes the spatial momenta and mass but not the temporal momenta. 
 The boundary condition obtained in (\ref{allow}) constrains the allowed modes and ensures that there is no particle current through the walls, thereby confining the fermion fields inside the slab-bag. The MIT Bag (B) boundary conditions are realized on the lattice as follows: 
\begin{equation}
\label{allowed}
     ap_1\rightarrow ap^{\text{B}}_1(n) = \left(n+\frac{1}{2}\right)\frac{\pi a}{d} = \left(n+\frac{1}{2}\right)\frac{\pi}{N}
\end{equation}
Here, $N = d/a$ is the lattice size in the compactified spatial direction. 
%It is necessary for these boundary conditions to be satisfied on the lattice, just as they are satisfied in the continuum for the confined massless free fermion field.  
%The expressions for the Casimir energy of the naive and Wilson fermion are calculated from this method analytically for ($1+1$)-dimensional spacetime, numerically for higher dimensions and plotted subsequently in Fig. \ref{mitbagplot1} \cite{Ishikawa:2020icy}. 
%This can also be considered a novel method to calculate the Dirac fermionic Casimir energy by taking the lattice spacing limit to zero.
The discretized momentum component under periodic (P) and antiperiodic (AP) boundary conditions, on the other hand, is:
\begin{equation}
\label{PAP}
ap_1 \rightarrow ap^{\text{P}}_1(n) = \frac{2n\pi}{N} \;\;;\;\; ap_1 \rightarrow ap^{\text{AP}}_1(n) = \frac{(2n+1)\pi}{N} 
\end{equation}
 Restriction of the momentum to the Brillouin zone bounds the integer $n$ for periodic, antiperiodic boundary conditions is, $0 \leq n^{\text{P,AP}} < N$, and MIT Bag boundary conditions is $0 \leq n^{\text{B}} < 2N$.
 %Unlike the continuum theory, both the zero point energies here are not divergent due to the lattice cut-off. 
 The Casimir energy on lattice is also calculated by subtracting the zero-point energy at finite lattice size $aE_0(N\rightarrow \infty)$ from the one at infinite lattice size $aE_0(N)$ \cite{Ishikawa:2020ezm}:
  %\begin{figure*}
%\centering
 %   \subfloat(a){{\includegraphics[width=8cm]{Casimir naive plots 1+1 Periodic.pdf} }} %
  %  \qquad
   % \subfloat(b){{\includegraphics[width=8cm ]{Casimir naive plots 1+1 Antiperiodic.pdf} }}%
   %  \subfloat(c){{\includegraphics[width=8cm]{Casimir Wilson plots 1+1 Periodic.pdf} }} %
    %\qquad
    %\subfloat(d){{\includegraphics[width=8cm]{Casimir Wilson plots 1+1 Antiperiodic.pdf} }} %
   % \subfloat(c){{\includegraphics[width=8cm]{Casimir naive %Plots 3 +1 Periodic.pdf}}}%
    %\qquad
    %\subfloat(d){{\includegraphics[width=8cm]{Casimir naive %Plots 3 +1 Antiperiodic.pdf}}}%
    %\caption{Massless and positively massive naive fermion.[(a),(b)] and [(c),(d)] represent the periodic and antiperiodic Casimir energy for naive and Wilson fermion 1+1-dimension respectively. The small windows represent the coefficient of Casimir Energy, which is constant in continuum and is useful for comparing the lattice and continuum theories. The dashed lines represent the leading order terms of $1/N$, obtained as an asyptotic form for the massless fermion in the large $N$-limit. }%
%\label{fig:test}
%\end{figure*}
\begin{align}
\label{def3+1}
aE^{\text{3+1}}_{\text{Cas}} &= aE_0(N) - aE_0(N \rightarrow \infty) \nonumber \\ 
              &= c_{\text{deg}}\int_{\text{BZ}} \frac{d^2ap_{\perp}}{(2\pi)^2}\left[-\sum_n aE(ap_{\perp},ap_1(n)) + N\int_{\text{BZ}}\frac{dap_1}{2\pi}aE(ap) \right] 
\end{align}
 The Casimir energy for the naive fermion is $2^D$ times the result for the Wilson fermion
%as can be seen from the results obtained on lattice in Fig. \ref{mitbagplot1}. 
due to the doubling multiplicity for the naive fermion. Here, $c_{\text{deg}}$ is the degeneracy factor which accounts for the spin of fermion and the naive fermion doubling.
%which will also be included in the $c_{\text{deg}}$ 
All comparisons are made taking care of this factor, so that the equivalence between analytic results of naive and Wilson can be established in (\ref{limnaive}) and (\ref{limwilson}) below.
%Casimir energy for  dimensions can be obtained accordingly. Note the change in dimension of the Brillouin zone. 
%Fermions have their characteristic negative sign. Due to a factor of (2) from the antiparticle degrees of freedom, the ($\frac{1}{2}$)  is dropped from the zero-point energy. 
 \subsection{Naive Fermions}
% We shall now discuss the Casimir energy $aE_{\text{Cas}}$, for naive lattice fermion ($nf$) in $(1+1)$- dimensional spacetime calculated from the definition. 
 %In momentum space, the Dirac operator for the naive fermion is :
%\begin{equation}
 %   aD_{nf}  = \eq i\sum_k \gamma_k \sin(ap_k) +am_f
%\end{equation}
The dispersion relation for naive fermions in ($D+1$)-dimensions obtained from (\ref{Drel}) is \cite{Gattringer:2010zz}:
\begin{equation}
\label{naived}
    a^2E_{nf}(ap) = \sum_{k=1}^{D}\sin^2(ap_k) +(am_f)^2
\end{equation}
As in Ref. \cite{Ishikawa:2020ezm}, the ($1+1$)-dimensions is treated analytically. 
%Numerical studies for the naive fermion in higher dimensions are also done 
By substituting (\ref{naived}) in the definition (\ref{def3+1}), the expression for Casimir energy of naive fermion with MIT Bag boundary conditions in ($D+1$)-dimensional spacetime is:
%\begin{equation}
%aE_{\text{Cas}}^{\text{3+1D,B,}nf} \equiv  aE_{\text{0}}^{\text{3+1D,B,}nf}(N) - %aE_{\text{0}}^{\text{3+1D,B,}nf}(N\rightarrow\infty)
%\end{equation}
 %\begin{equation}
 %= c_{\text{deg}}\int \frac{d^2ap_{\perp}}{(2\pi)^2}\left[
 %\begin{aligned}
 %-\sum_n \sqrt{\sin^2 \frac{(n+1/2)\pi}{N}+ \sum_{k=2,3}\sin^2  &+ (am_f)^2} + \\& %N\int_{\text{BZ}}\frac{dap_1}{2\pi}\sqrt{ \sum_{k=1,2,3}\sin^2 ap_k+ (am_f)^2}  
 %\end{aligned}
 %\right]
 %\end{equation}
 \begin{align}
    \label{naivemitbagexp}
    &aE_{\text{Cas}}^{\text{D+1,B,}nf} \equiv  aE_{\text{0}}^{\text{D+1,B,}nf}(N) - aE_{\text{0}}^{\text{D+1,B,}nf}(N\rightarrow\infty)\\
&= c_{\text{deg}}\int_{\text{BZ}} \frac{d^{D-1}ap_{\perp}}{(2\pi)^2}\Bigg[
-\sum_n \sqrt{\sin^2 \frac{(n+1/2)\pi}{N}+ \sum_{k=2}^{D}\sin^2 (ap_k)  + (am_f)^2}\;\; + \\&\;\;\;\;\;\;\;\;\;\;\;\;  \;\;\;\;\;\;  \;\;\;\;\;\;  \;\;\;\;\;\;  \;\;\;\;\;\;  \;\;\;\;\;\;     N\int_{\text{BZ}}\frac{dap_1}{2\pi}\sqrt{ \sum_{k=1}^{D}\sin^2 (ap_k)+ (am_f)^2}\Bigg] \nonumber
\end{align} 
%Similarly, the expression for ($2+1$)-dimensions is used for numerical analysis. 
%The derivations of Casimir energy in ($1+1$)-dimensional spacetime using the MITBag (B) momenta boundary conditions for the massless naive fermion are presented in Appendix B (\ref{naivemitbag}) using the Abel- Plana formulae for finite range. 
In the case for $D=1$, the series expansions are substituted, and the exact expression is calculated using the Abel-Plana formulae for finite range \cite{Saharian:2006iv,  Saharian:2007ph}, to be:
\begin{align}
     aE_{\text{Cas}}^{\text{1+1,B,}nf} & =\frac{N}{\pi} -\frac{1}{2}\csc(\frac{\pi}{2N})\label{mitbagnaive}\\
   & =-4\left[-\frac{d}{4a\pi} + \frac{1}{8}\left\{\frac{2d}{a\pi} + \frac{1}{6}\frac{a\pi}{2d} +\frac{7}{360}\left(\frac{a\pi}{2d}\right)^3 + ...\right\} \right]\nonumber\\
  \Rightarrow E_{\text{Cas}}^{\text{1+1,B,}nf} & = -\frac{\pi}{24d} - \frac{7\pi^3a^2}{5760d^3} + \mathcal{O}(a^4) 
 \end{align} 
 Therefore, the Casimir energy obtained per unit area in the continuum limit $a\rightarrow 0$, is:
\begin{equation}
\label{limnaive}
\lim_{a \to 0}E_{\text{Cas}}^{\text{1+1,B},nf} = -\frac{\pi}{24 d}
\end{equation}
As expected, taking the naive fermion doubling correction in one spatial dimension into account, the above result in continuum limit is equivalent to the expression obtained for Dirac fermion (\ref{fercont}).  These expressions for Casimir energy of naive fermions, without the doubling factor are plotted in Fig. \ref{mitbagplot1}. Studies for the naive fermion for $D>1$ are done numerically. Results for (2+1)- and (3+1)- dimensions have been verified to be exactly equal to the continuum expressions. Only results for (3+1)-dimensions have been shown in Fig. \ref{mitbagplot1}.
 \subsection{Wilson Fermion}
  We shall now discuss the Casimir energy for Wilson lattice fermion in $(1+1)$-dimensional spacetime.
  %The Dirac operator obtained for Wilson fermion with fermion mass $m_f$ and the Wilson parameter $r$ in momentum space is:
%\begin{equation}
%\label{Wilsonoperator}
%    aD_{\text{W}}  =  i\sum_k \gamma_k \sin(ap_k) + r\sum_k (1- \cos(ap_k) + am_f
%\end{equation}
The Wilson term proportional to $r$ in the Dirac operator breaks the chiral symmetry and acts as a momentum dependent mass term introduced to eliminate fermion doubling. 
%The dispersion relation for Wilson fermion is 
%\begin{equation}
%    \label{Wilsonrelation}
 %   a^2E^2_{\text{W}}(ap) = \sum_k \sin^2(ap_k) +\left[r\sum_k(1-\cos(ap_k)+am_f\right]^2
%\end{equation}
The expression for Casimir energy of Wilson fermion in ($D+1$) is similar to the one for naive fermion, where the Wilson Dispersion relation, obtained using (\ref{Drel}), is substituted in (\ref{def3+1}). 
%The detailed derivations of Casimir energy in $(1+1)$-dimensional spacetime for MITBag (B) boundary conditions using the Abel-Plana formulae for finite range are presented in Appendix B (\ref{Wilsonmitbag}). 
%Series expansions are substituted, and t
Using the Abel-Plana formulae for finite range like in the naive fermion case, setting $r=1$, the following expressions are obtained:
\begin{align}
    aE_{\text{Cas}}^{\text{1+1,B,W}}& =\frac{4N}{\pi} -\csc(\frac{\pi}{4N})\label{mitbagwilson} \\
   & =-4
   \left[-\frac{d}{a\pi} + \frac{1}{4}\left\{\frac{4d}{a\pi} + \frac{1}{6}\frac{a\pi}{4d} +\frac{7}{360}\left(\frac{a\pi}{4d}\right)^3 + ...\right\} \right]\nonumber\\
 \Rightarrow E_{\text{Cas}}^{\text{1+1,B,W}} & = -\frac{\pi}{24d} - \frac{7\pi^3a^2}{23040d^3} + \mathcal{O}(a^4)
 \end{align}
    \begin{figure}[t!]
\centering
    \footnotesize{(a)}{{\includegraphics[width=6.25cm]{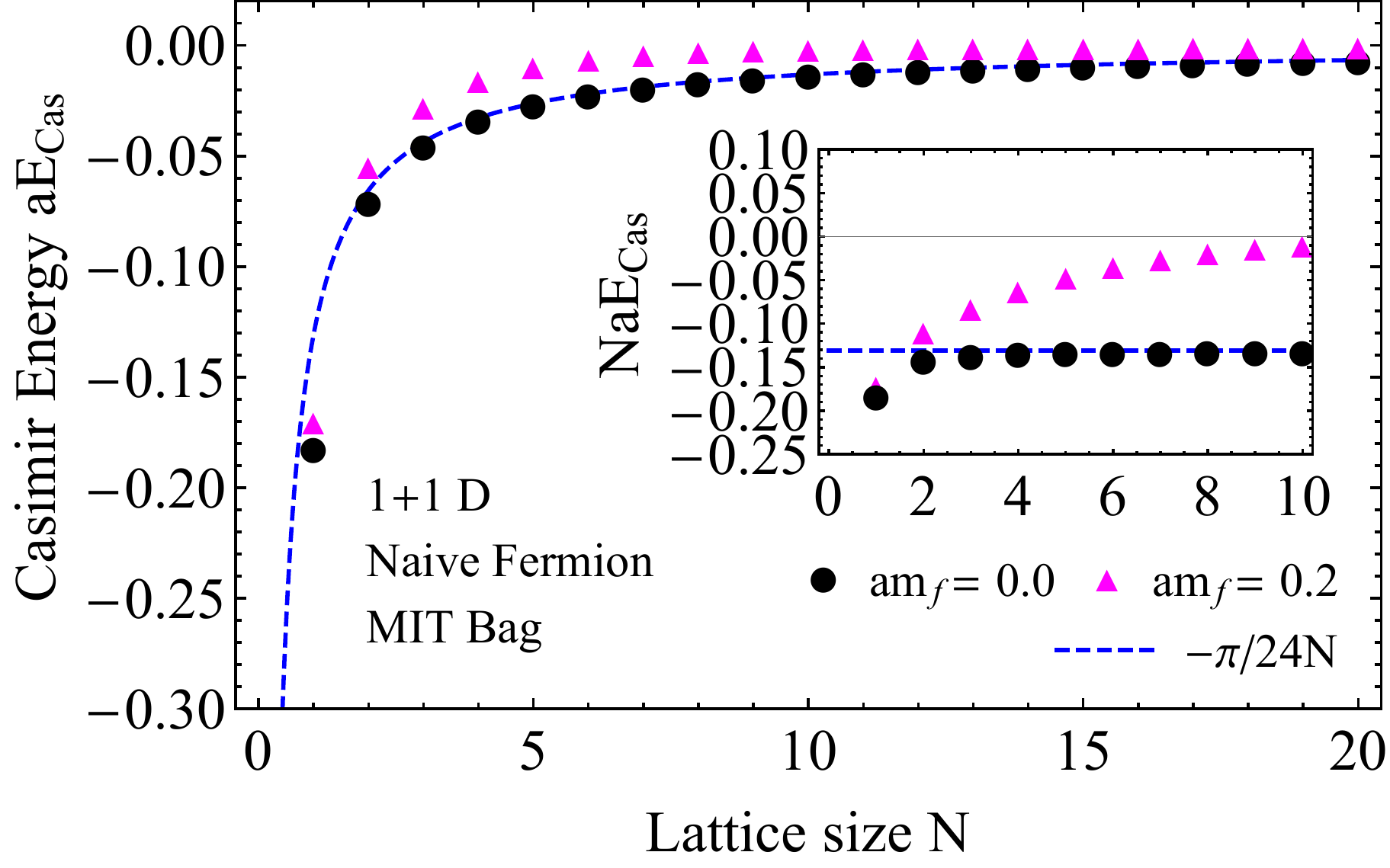} }} %
    \footnotesize{(b)}{{\includegraphics[width=6.25cm ]{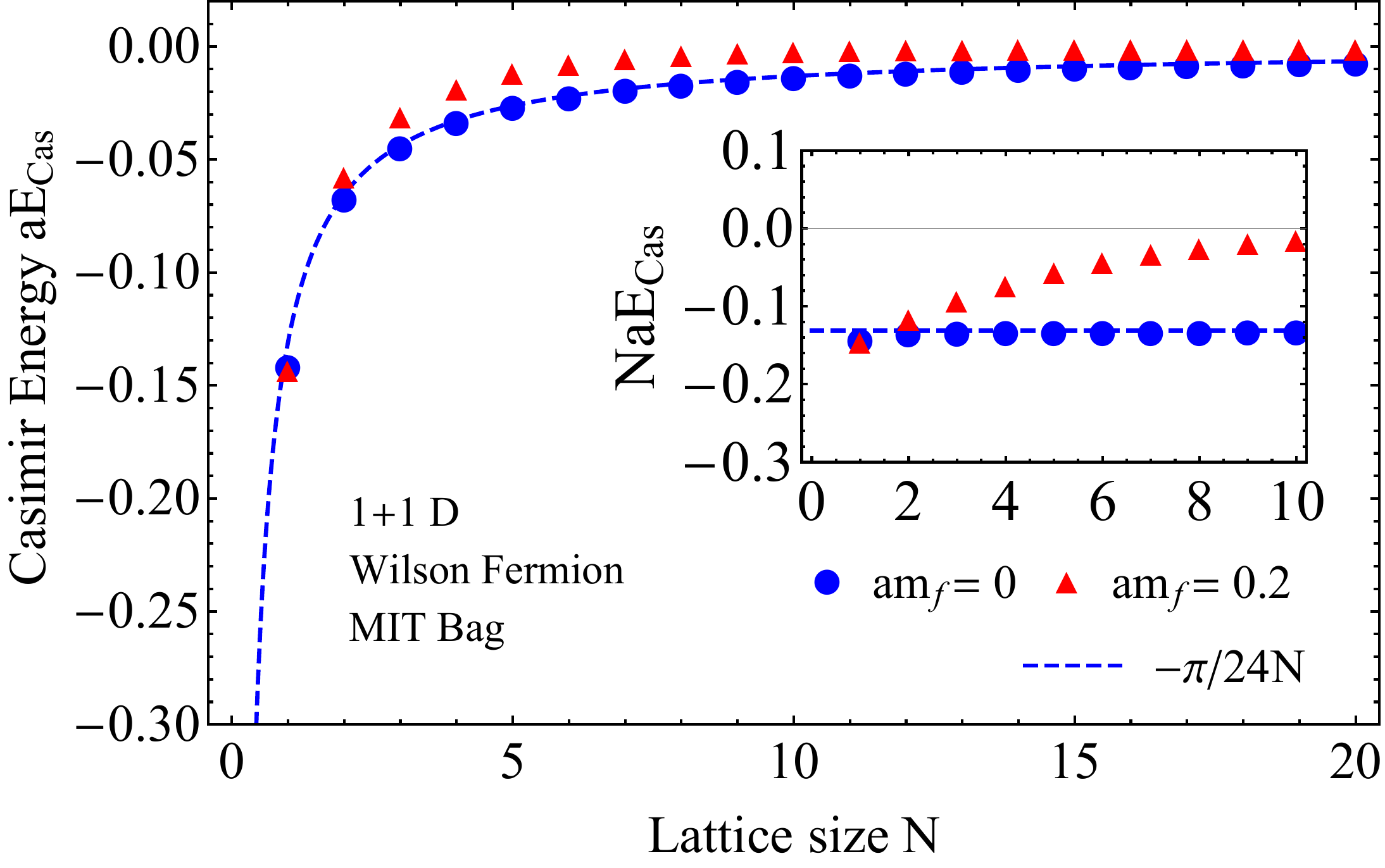} }}%
  %  \subfloat\footnotesize{(c)}{{\includegraphics[width=6.25cm]{Casimir naive plots 2+1 mitbag.pdf}}} 
   % \subfloat\footnotesize{(d)}{{\includegraphics[width=6.25cm]{Casimir Wilson plots 2+1 mitbag.pdf}}}
   \footnotesize{(c)}{{\includegraphics[width=6.25cm]{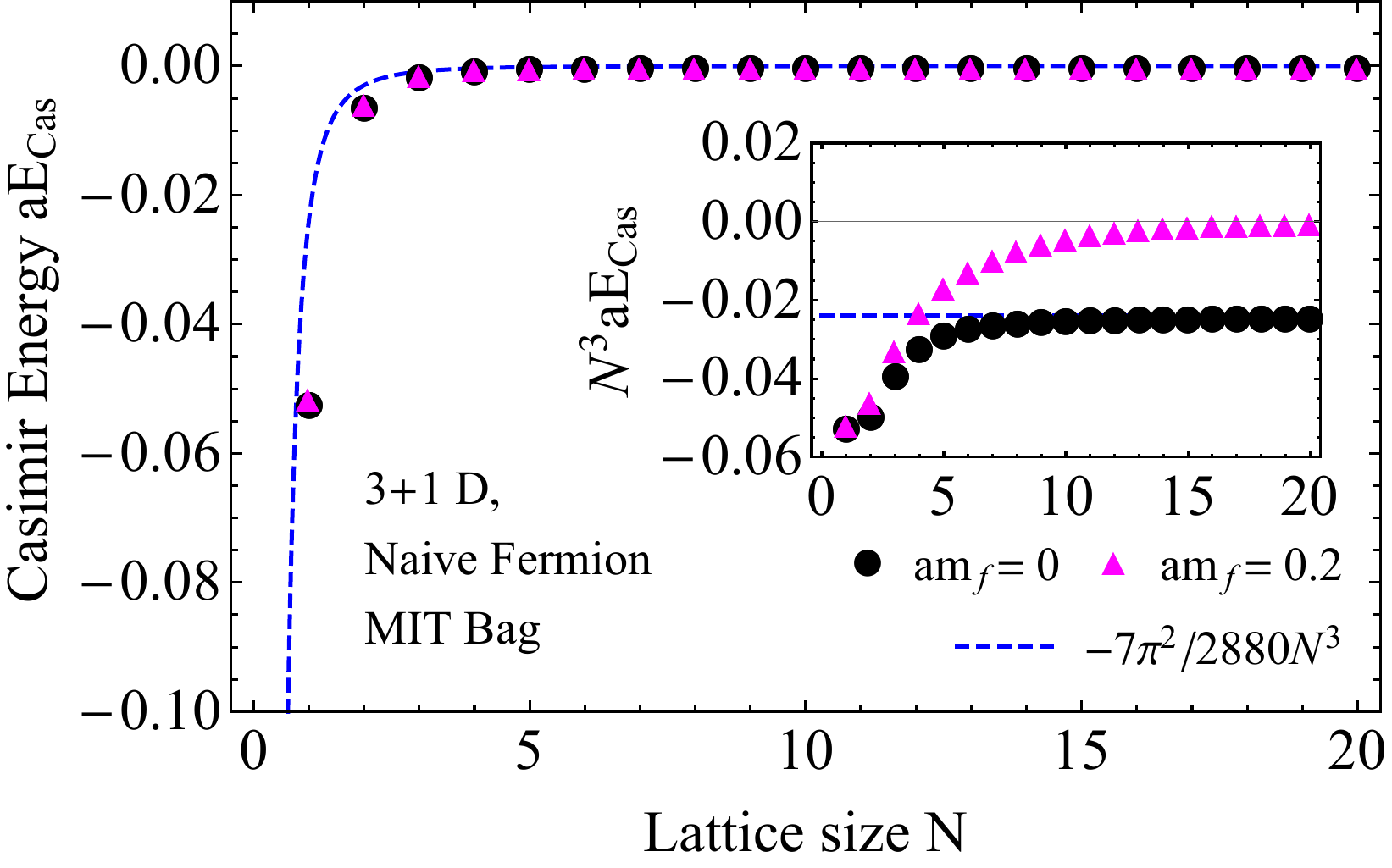}}} %
    \footnotesize{(d)}{{\includegraphics[width=6.25cm]{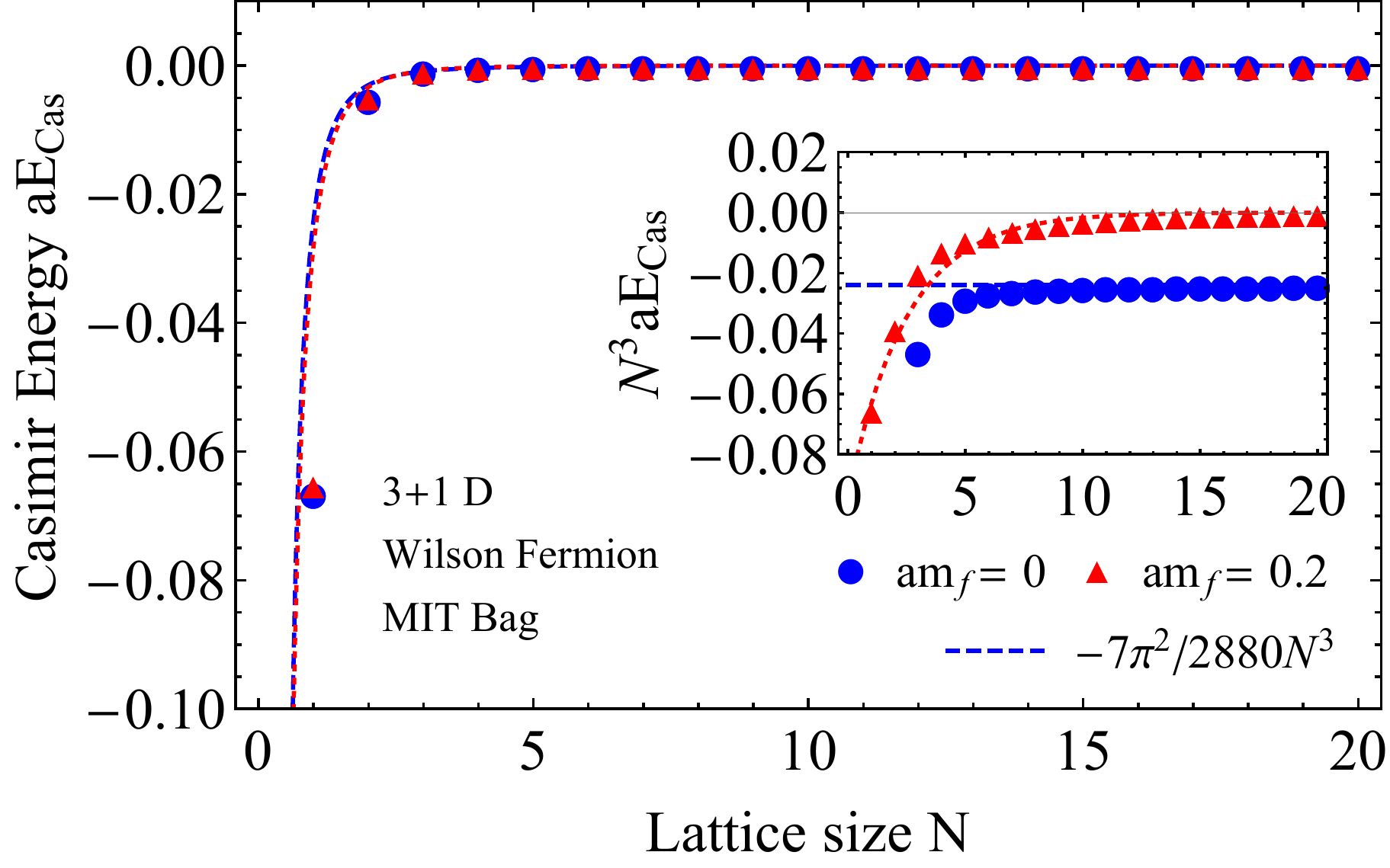}}}%
   % \subfloat(c){{\includegraphics[width=8cm]{Casimir naive %Plots 3 +1 Periodic.pdf}}}%
    %\qquad
    %\subfloat(d){{\includegraphics[width=8cm]{Casimir naive %Plots 3 +1 Antiperiodic.pdf}}}%
    \caption{{Casimir energy for massless and positively massive Wilson and naive fermion with MIT Bag boundary conditions. [(a),(b)] and [(c),(d)] represent the MIT Bag Casimir energy for naive and Wilson fermion in ($1+1$)- and ($3+1$)-dimension respectively. Subfigure (d) also confirms the exponential decay of Casimir energy for massive fermions with $N$ on a lattice.}}%
\label{mitbagplot1}
\end{figure}
Therefore, In the limit $a\rightarrow 0$, the Casimir energy obtained per unit area is:
\begin{equation}
\label{limwilson}
\lim_{a \to 0}E_{\text{Cas}}^{\text{1+1,B,W}} = -\frac{\pi}{24 d}
\end{equation}
%The result obtained for the massless Wilson and naive fermion for the allowed frequency stated in (\ref{bound}) is obtained and found to be different.
%The expressions for the Casimir energy of the naive and Wilson fermion are plotted using the MITBag (B) boundary conditions.
The exact expressions obtained for the massless naive and Wilson fermion using MIT Bag boundary conditions in $(1+1)$-dimension are (\ref{mitbagnaive}) and (\ref{mitbagwilson}) respectively.
%\begin{align}
 %\label{obt}
 %aE^{\text{1+1D,B,}nf}_{\text{Cas}} &= \frac{2N}{\pi}-\csc(\frac{\pi}{2N})\label{naiveo}\\aE^{\text{1+1D,{B},W}}_{\text{Cas}} &= \frac{4N}{\pi}-\csc(\frac{\pi}{4N})\label{wilsono}
 %\end{align}
   The leading terms of the order $1/N$ in the zero lattice spacing limit of these analytic results and the numerical results for naive lattice fermion in ($3+1$)- dimensions are plotted in Fig. \ref{mitbagplot1}. Note that the exponential suppression of Casimir energy for massive fermionic fields, which is derived for the fermion mass limit $md\gg1$ in (\ref{massivefer}), is also verified in numerical calculations on a lattice as shown in Fig. \ref{mitbagplot1}(d) for comparison. 
   
   In $(D+1)$-dimensional spacetime, the smaller windows in each plot represent the coefficient of Casimir energy, $N^{D}\cdot a\cdot E_{\text{Cas}}$, which is constant in continuum and thus is used to observe the precision of agreement between the lattice and continuum results. 
 %The suppression of Casimir energy for massive naive fermion, as compared to the massless case can also be verified from Fig. \ref{dispersionrelations}. 
%The coefficient $\frac{1}{2}$ in the definition (\ref{def3+1}) involving the zero-point energy is cancelled by this factor. 
%The fermion field $\psi$ has $N_D$ components given by $2^{\frac{D+1}{2}}$ if $D$ is odd and by $2^{\frac{D}{2}}$ if $D$ is even.
The result obtained for the Wilson fermion 
%in ($1+1$)- and ($2+1$)-dimension
matches with the continuum result (\ref{fercont}) exactly.  %Although, in the ($3+1$)-dimension, we have to take into account that the degeneracy factor $c_{\text{deg}} = 2$ arising from the spin degrees of freedom. Thus, in this case, the overall factor is four (as $N_D$ = 4). Once this additional factor of two is taken care of in the numerical results obtained on the lattice, the result we obtained for the Wilson fermion for ($3+1$)-dimension matches exactly with the continuum result in (\ref{cont}). 
%In the ($1+1$)- and ($2+1$)-dimensional case, the expression for naive fermion is twice and four times the continuum result, respectively, due to the fermion doubling in spatial directions. Also remember that the degeneracy factor $c_{\text{deg}} = 1$ in these cases, and only a factor of two 
%($N_{{D}} = 2$ for both cases)
%accounting for the particle-antiparticle degeneracy is considered. 
%In this case, the result for the naive fermion is eight times the expression for the Wilson fermion. Thus, 
 \begin{figure}[t!]
\centering
    \footnotesize{(a)}{{\includegraphics[width=6.25cm]{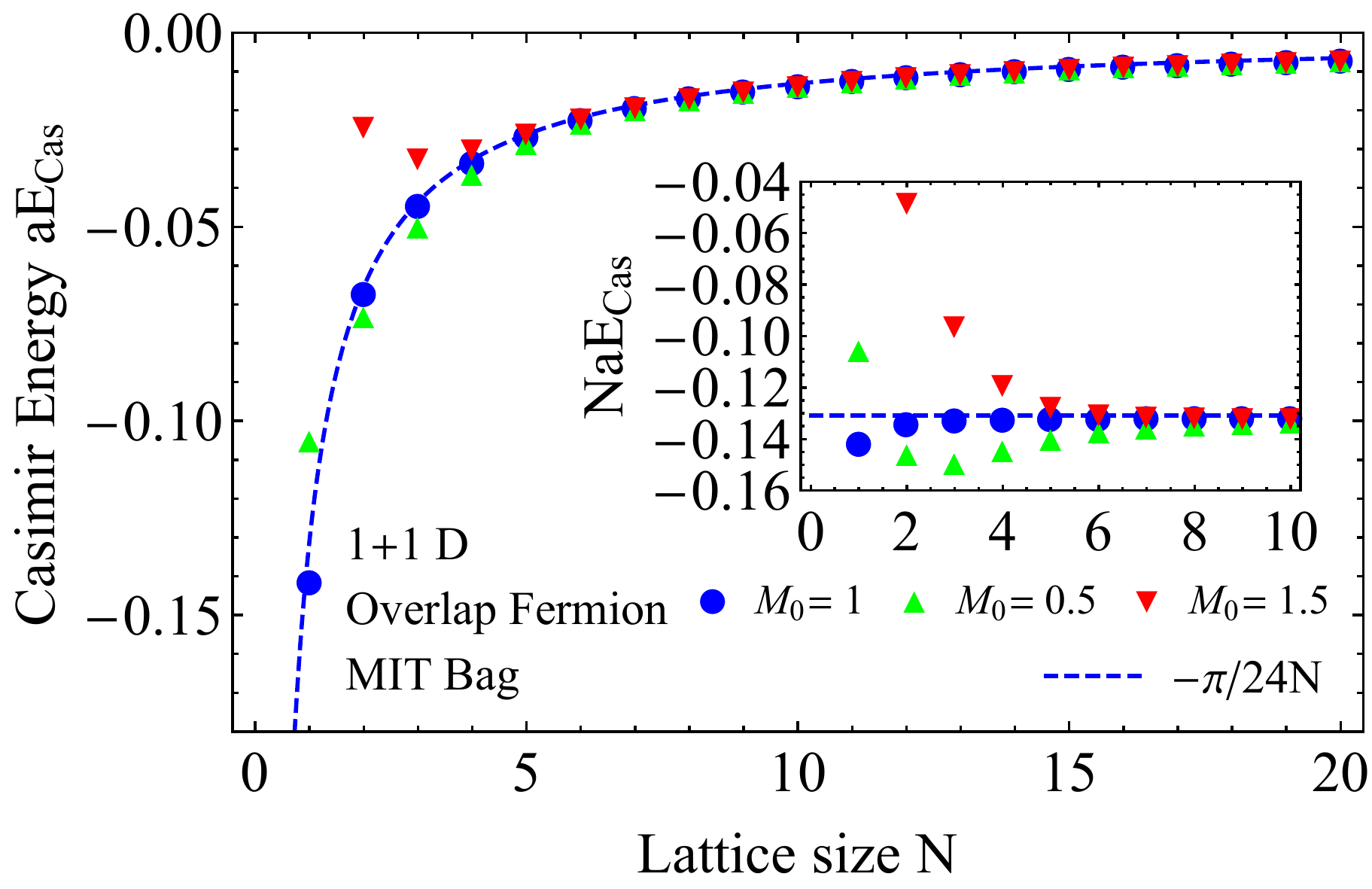} }} 
   % \subfloat\footnotesize{(b)}{{\includegraphics[width=6.25cm ]{Overlap fermions 2+1 plots mitbag.pdf} }}%
    \footnotesize{(c)}{{\includegraphics[width=6.25cm]{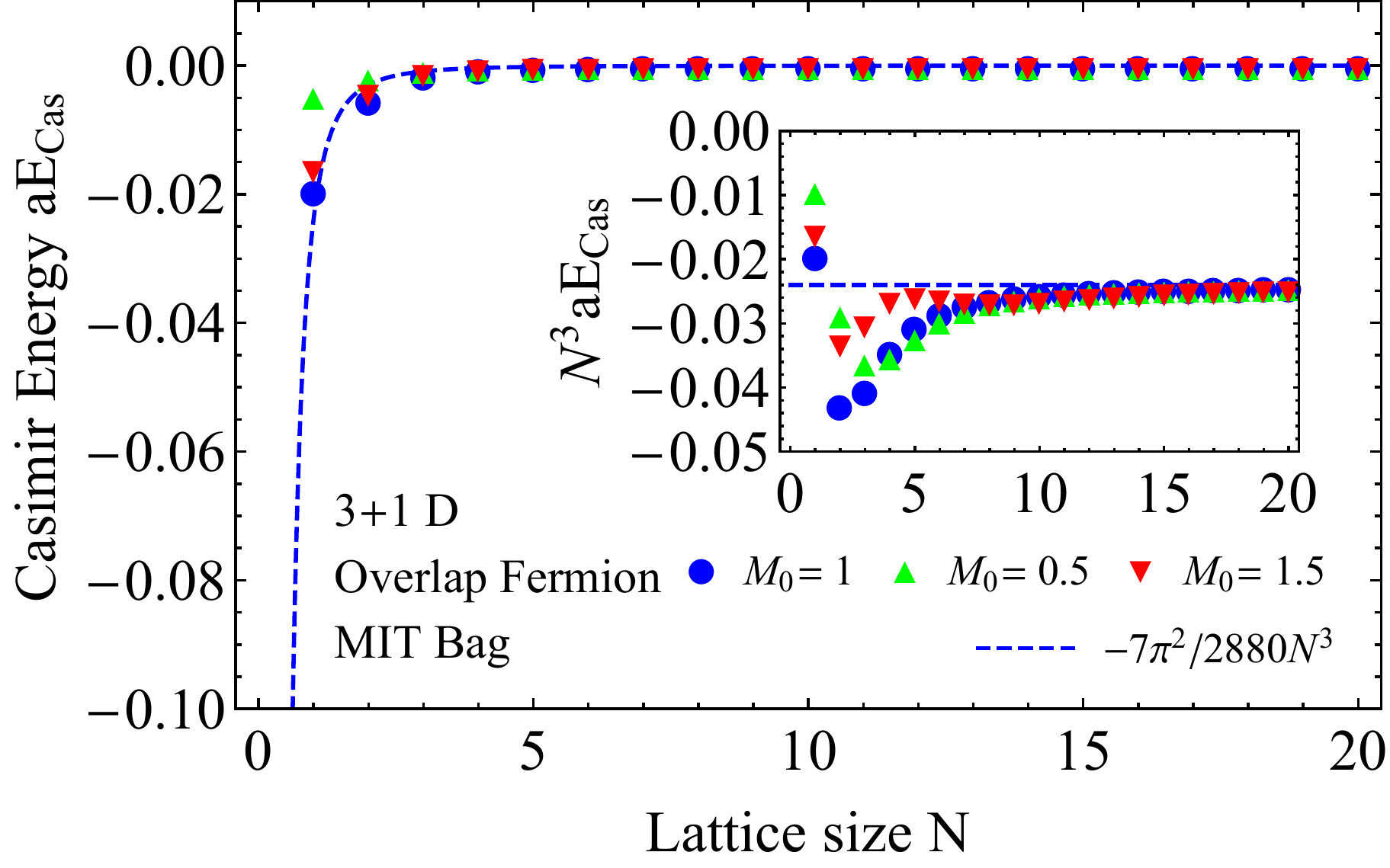}}}
   % \subfloat(c){{\includegraphics[width=8cm]{Casimir naive %Plots 3 +1 Periodic.pdf}}}%
    %\qquad
    %\subfloat(d){{\includegraphics[width=8cm]{Casimir naive %Plots 3 +1 Antiperiodic.pdf}}}%
    \caption{\footnotesize{Casimir energy for overlap fermion 
    %with \textcolor{red}{MDW} kernel
     with MIT Bag boundary conditions. (a),(b) and (c) represent the Casimir energy in ($1+1$)- and ($3+1$)-dimension respectively. }}%
\label{omitbag}
\end{figure}
\subsection{Overlap Fermion}
\label{overlapmitbag}
The analytic calculation of the Casimir effect for overlap fermions in (1+1)-dimensions turned out to be intractable. Thus, the results for the overlap fermion are numerically calculated for all dimensions, using the MIT Bag boundary conditions by substituting the corresponding Dirac operator into the definition (\ref{def3+1}). 
%The fermion mass ($am_f$) is now replaced by the the parameter $
%am_f \rightarrow 
%-M_0$  and varied as $M_0 = 0.5$, $1.0$ and $1.5$.
The Dirac operator for overlap fermion is defined as:
%in terms of the \textcolor{red}{M\"{o}bius} domain wall (\textcolor{red}{MDW}) kernel operator $D_{\text{\textcolor{red}{MDW}}}$ as:
\begin{equation}
    a{\mathcal{D}}_{\text{OV}}\equiv2 M_0 \times\frac{(1+am_f)+(1-am_f)V}{2}
\end{equation}
where $am_f$ is the fermion mass
%, $m_{\text{PV}}$ is the Pauli-Villars mass
 and $V$ is defined as:
\begin{equation}
\label{V}
    V\equiv\gamma_5\;\text{sign}(\gamma_5 a {\mathcal{D}}_{\text{W}}) = \frac{{\mathcal{D}}_{\text{W}}}{\sqrt{{\mathcal{D}}_{\text{W}}^\dagger {\mathcal{D}}_{\text{W}}}}
    %\text{ and } aD_{\text{W}}\equiv \frac{b(aD_{\text{W}})}{2+c(aD_{\text{W}})}
\end{equation}
%and the \textcolor{red}{M\"{o}bius} domain wall (\textcolor{red}{MDW}) kernel operator is:
%\begin{equation}
%\label{\textcolor{red}{MDW}}
 %   aD_{\text{\textcolor{red}{MDW}}}\equiv \frac{b(aD_{\text{W}})}{2+c(aD_{\text{W}})}
%\end{equation}
with 
%the $b$ and $c$ are called \textcolor{red}{M\"{o}bius} parameters, and 
 ${\mathcal{D}}_{\text{W}}$ as the previously defined Wilson Dirac operator with $r=1$ and a negative mass parameter -$M_0$. We used $M_0 = 0.5$, $1.0$ and $1.5$. 
%Note that this study can also be done with the domain wall fermion kernel with a finite extra dimension. 
%In the case where $c=0$ $m_{\text{PV}}=1.0$ are kept as fixed parameters in the dispersion relation throughout our calculation. 
The results obtained for $(1+1)$- and $(3+1)$-dimensions are plotted in Fig. \ref{omitbag}.
It is easy to notice in ($1+1$)- dimensions that the overlap fermion $M_0 = 1.0$ case and massless ($am_f = 0$) Wilson fermion case have equivalent dispersion relations. The expressions obtained numerically for overlap fermions 
%with \textcolor{red}{MDW} kernel 
in higher dimensions, also match precisely with the naive and Wilson fermions in the zero lattice spacing limit $a\rightarrow0$ and, subsequently, also with the continuum result.
\section{Naive fermion and Series Extrapolation in large $N$-Limit}
\label{extrapol}
\begin{figure}[t!]
\centering
  \footnotesize{(a)}{{\includegraphics[width=6.25cm]{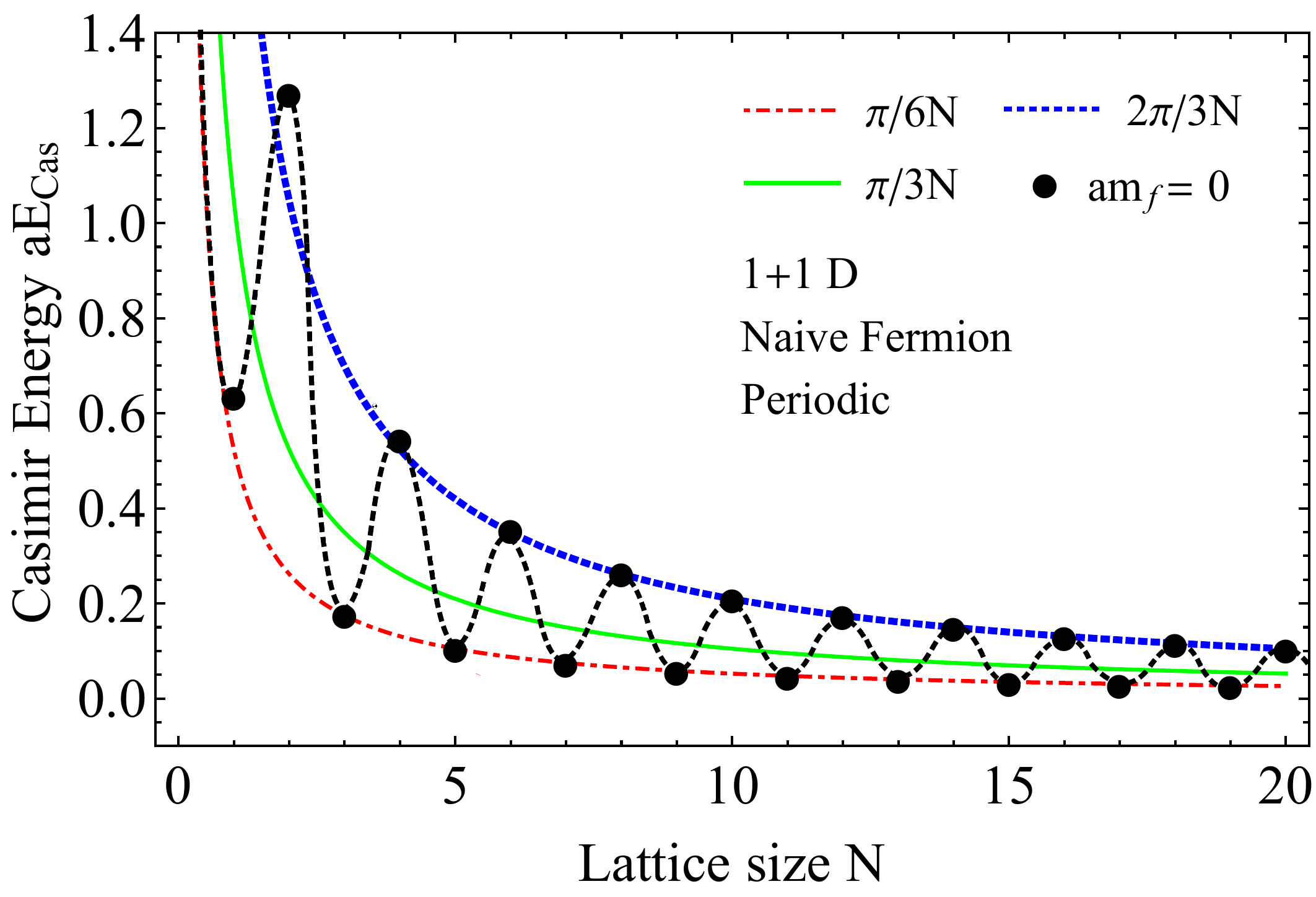} }} 
   \footnotesize{(b)}{{\includegraphics[width=6.25cm ]{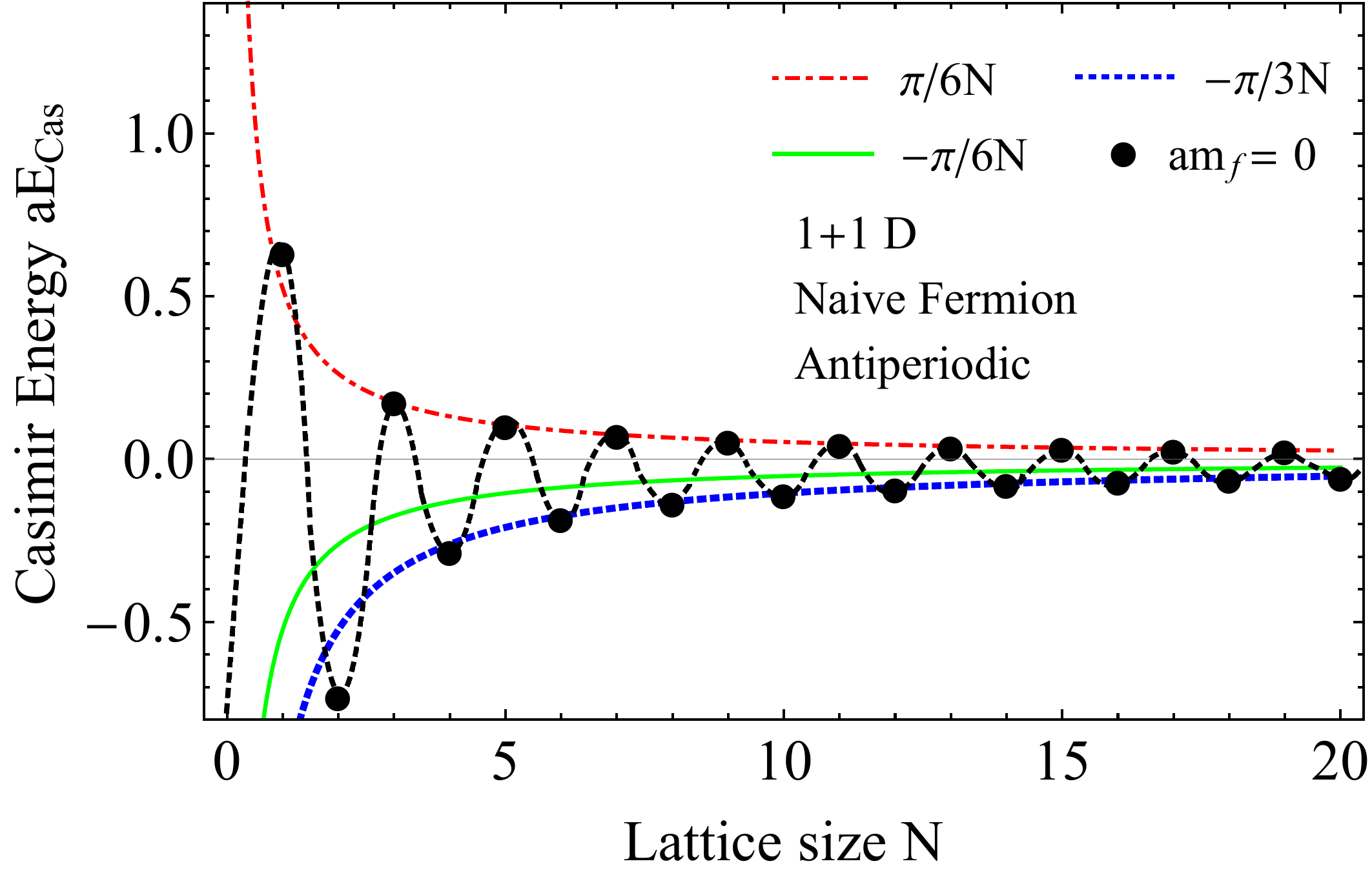} }}%
\caption{\footnotesize{Oscillation of Casimir energy for positively massive and massless naive fermion for odd and even $N$. [(a),(b)] represent the periodic and antiperiodic boundary conditions  in ($1+1$)-dimension respectively.}}%
\label{osci}
\end{figure}

%\subsection{For naive fermion}
%The expression for the Casimir energy of naive fermion with periodic (P) boundary conditions, used to numerically study the Casimir energy in higher dimensions is:
%   \begin{align}
 %  \label{naiveexp}
  %     &aE_{\text{Cas}}^{\text{3+1D,P,}nf} \equiv  aE_{\text{0}}^{\text{3+1D,P,}nf}(N) - aE_{\text{0}}^{\text{3+1D,P,}nf}(N\rightarrow\infty)\\
   %    &=  c_{\text{deg}}\int \frac{d^2ap_{\perp}}{(2\pi)^2}\Bigg[-\sum_{n=0}^{N-1} \sqrt{\sin^2 \frac{2n\pi}{N}+\sum_{k=2,3}\sin^2(ap_k) + (am_f)^2} \\& \;\;\;\;\;\;\;\;\;\;\;\;\;\;\;\;\;\;\;\;\;\;\;\;\;\;\;\;\;\;\;\;\;\;\;\;\;\;\;\;\;\;\;\;\;\;\;\;\;\;\;\;\;\;\;\;\;\;\;\;+ N\int_{\text{BZ}}\frac{dap_1}{2\pi}\sqrt{ \sum_{k=1,2,3}\sin^2(ap_k) + (am_f)^2}  \Bigg]\nonumber
%  \end{align}
While it is reassuring that one obtains the {\em same} results for the naive, Wilson and overlap fermions for the physically appealing MIT bag boundary conditions, as shown in the section above, it still is a bit disturbing that such validation of universality is not seen in the results of Ref. \cite{Ishikawa:2020ezm} which employed periodic and antiperiodic boundary conditions.   We therefore now turn to examine the naive fermion case with those boundary conditions in an attempt to shed more light on this problem.  The exact expressions for Casimir energy $aE_{\text{Cas}}$ of massless naive fermion on lattice in ($1+1$)-dimensional spacetime with periodic and antiperiodic boundary conditions, were first obtained in Ref \cite{Ishikawa:2020ezm, Ishikawa:2020icy}. 
%For detailed derivation of these expressions, refer to Ref. \cite{Ishikawa:2020icy}. 
%They are as follows for periodic (P) boundary conditions:
%\begin{equation} 
%aE^{\text{1+1D,P},nf}_{\text{Cas}} = \begin{dcases}
%                \frac{2N}{\pi}- 2\cot(\frac{\pi}{N}) & \text{if $N=\text{even}$}\\ 
%                \frac{2N}{\pi}- \cot(\frac{\pi}{2N})& \text{if $N=\text{odd}$}\\
%               \end{dcases}  
%               \label{naivep}
%\end{equation}
Their results 
%in the continuum limit for odd and even lattice sizes, individually 
are:
\begin{equation}
\label{naivePcont}
\lim_{a \to 0}E_{\text{Cas}}^{\text{1+1,P,}nf} = \frac{\pi}{6d}\;\;\;\;\;(\text{odd } N)\;\;;\;\;\lim_{a \to 0}E_{\text{Cas}}^{\text{1+1,P,}nf} = \frac{2\pi}{3d}\;\;\;\;\;(\text{even } N)
\end{equation}
 for periodic boundary. 
 %the expressions for antiperiodic (AP) boundary conditions are:
%\begin{equation}
% aE_{\text{Cas}}^{\text{1+1D,AP},nf} =   \begin{dcases}
 %               \frac{2N}{\pi}- 2\csc(\frac{\pi}{N}) & \text{ if } N=\text{even}\\ 
 %               \frac{2N}{\pi}- \cot(\frac{\pi}{2N}) & \text{ if } N=\text{odd}
  %             \end{dcases} 
   %            \label{naivea}
%\end{equation}
%The continuum limit ($a\rightarrow0)$ expressions for the original Dirac fermion are also obtained subsequently as:
 Similarly, for antiperiodic boundary conditions, the result is:
\begin{equation}
\label{naiveAPcont}
\lim_{a \to 0}E_{\text{Cas}}^{\text{1+1,AP,}nf} = \frac{\pi}{6d}\;\;\;\;\;(\text{odd } N)\;\;;\;\;\lim_{a \to 0}E_{\text{Cas}}^{\text{1+1,AP,}nf} = -\frac{\pi}{3d}\;\;\;\;\;(\text{even } N)
\end{equation}
The continuum Casimir energy for massless Dirac fermion using periodic and antiperiodic conditions in (\ref{PAP}) is found to be:
 \begin{equation}
 \label{cont}
     E^{1+1,\text{cont},\text{P}}_{\text{Cas}} = \frac{\pi}{3d}\;;\;E^{1+1,\text{cont},\text{AP}}_{\text{Cas}} = -\frac{\pi}{6d}
 \end{equation}
Thus, the Casimir energy depends on the type of boundary conditions. 
%The Wilson and overlap fermions  reproduce the continuum results in the $ a \to 0$ limit.
Comparing (\ref{fercont}) and (\ref{cont}) one observes a peculiar apparent violation of universality as first shown in Ref \cite{Ishikawa:2020ezm}. The results depend on the whether the lattice size is odd or even, and their respective continuum limits do not agree with (\ref{cont}). 
%Fig. \ref{} displays the results from (\ref{})-(\ref{}).
Since $d= N \times a$, where $N$ now can be even or odd, $aE \propto 1/N$ with differing constants for the continuum result as well as odd/even lattice results, as seen in   
(\ref{naivePcont}-\ref{cont}).   These curves are shown with appropriate labels in the panels of Fig. \ref{osci} for both periodic and antiperiodic boundary conditions.

%Note that for naive fermion, one encounters different expressions for Casimir energy on odd and even lattice sizes in the continuum limit.
In the limit of large lattice size $N$, one can compute the difference between the Casimir energy for odd and even lattice sizes from Ref. \cite{Ishikawa:2020ezm}, as
\begin{equation}
\label{diff}
    aE^{\text{1+1},nf}_{\text{Cas}}(\text{odd } N)-aE^{\text{1+1},nf}_{\text{Cas}}(\text{even } N) = \mp4\tan(\frac{\pi}{4N})
\end{equation} for periodic and antiperiodic boundary conditions respectively, which is an oscillating yet rapidly converging function. %In case of no oscillation of Casimir energy with the lattice size, this difference will ideally be zero. 
The oscillations as $N$ increases from odd to even are depicted 
by the black dotted function in Fig. \ref{osci} in both the panels.  The continuum expressions are represented by the continuous green line in Fig. \ref{osci} and are seen to average out the oscillations well.
This observation provided us a hint to look for suitable methods to understand the behaviour of Casimir energy for naive fermion as part of a single series. %in large lattice sizes $N$.
%The techniques of series acceleration are often applied in numerical analysis to improve the speed of numerical convergence.
In (\ref{def3+1}), the Casimir energy is expressed as a difference between the integral and summation of the same function in one spatial dimension on the lattice. It is the sum-part which contributes to the rapidly oscillating behaviour above.
The Euler-Maclaurin formula\footnote{The routine attribute of numerical summation in Wolfram Mathematica, $\texttt{"NSum"}$ uses the Euler-Maclaurin formulae using $\texttt{"Method}\rightarrow\texttt{EulerMaclaurin"}$ to extrapolate the series after a specified number of terms in the summation using the $\texttt{"NSumTerms"}$ attribute. %The expression used for this is mentioned in Appendix (\ref{EMMathematica}).
}  
%Wynn's epsilon ($\epsilon$) method and Richardson extrapolation \cite{osada, Mark}
is a series acceleration technique often applied 
%numerical analysis
%to improve the speed of convergence of such sums
in such cases,
%The Euler-Maclaurin method is employed in cases
where the last term of the series tends to 0 as $n\rightarrow\infty$, and expresses the Euler-Maclaurin sum of a function as an infinite series in terms of integral of the same function and higher-order derivative differences.
%This suggests that the Euler-Maclaurin method to be the best suited for our problem. %{and is therefore employed}.
%{\bf Yash this needs to be better explained.}
     \begin{figure}[t!]
\centering
    \footnotesize{(a)}{{\includegraphics[width=6.25cm]{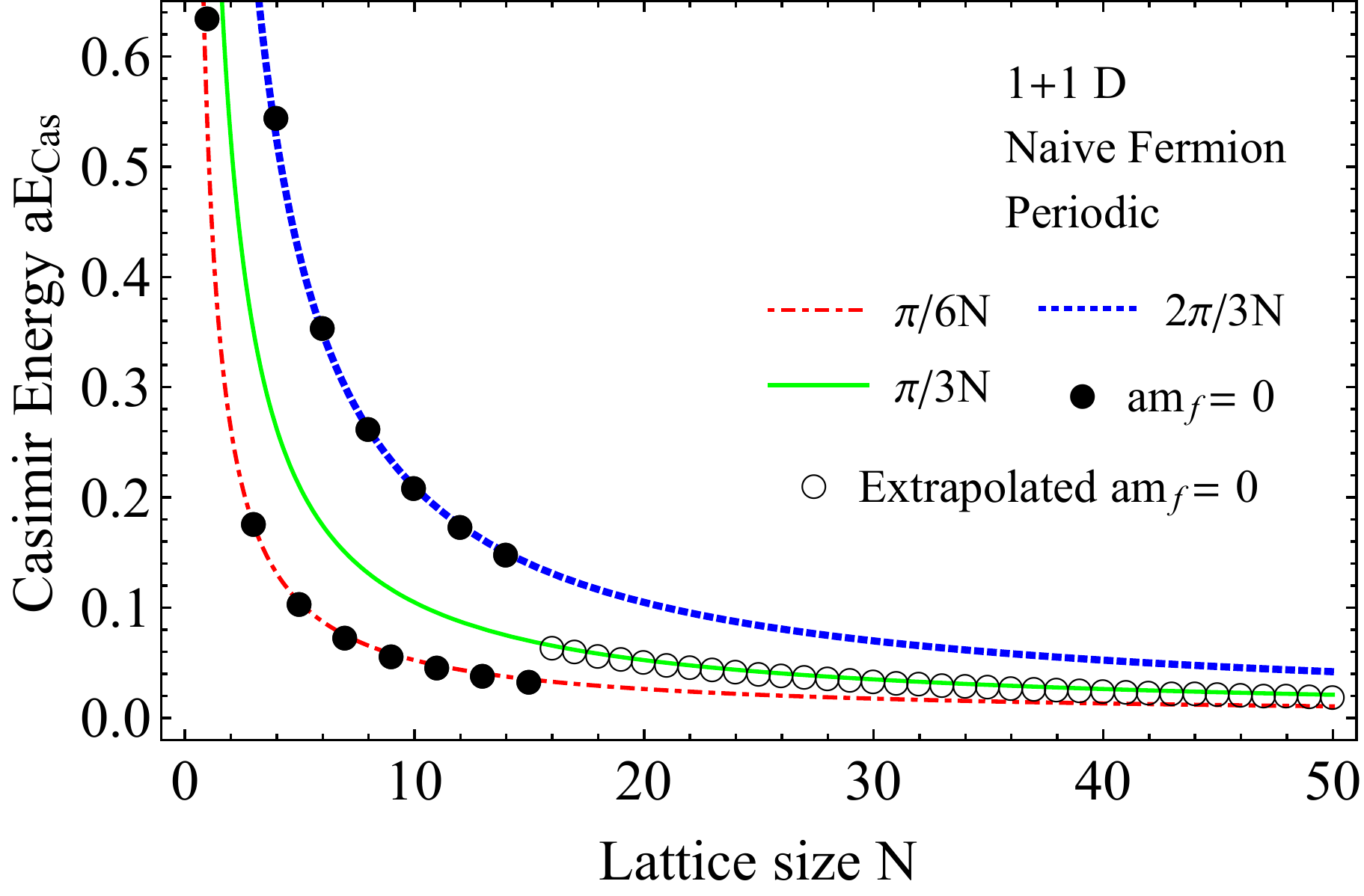} }} %
  \footnotesize{(b)}{{\includegraphics[width=6.25cm ]{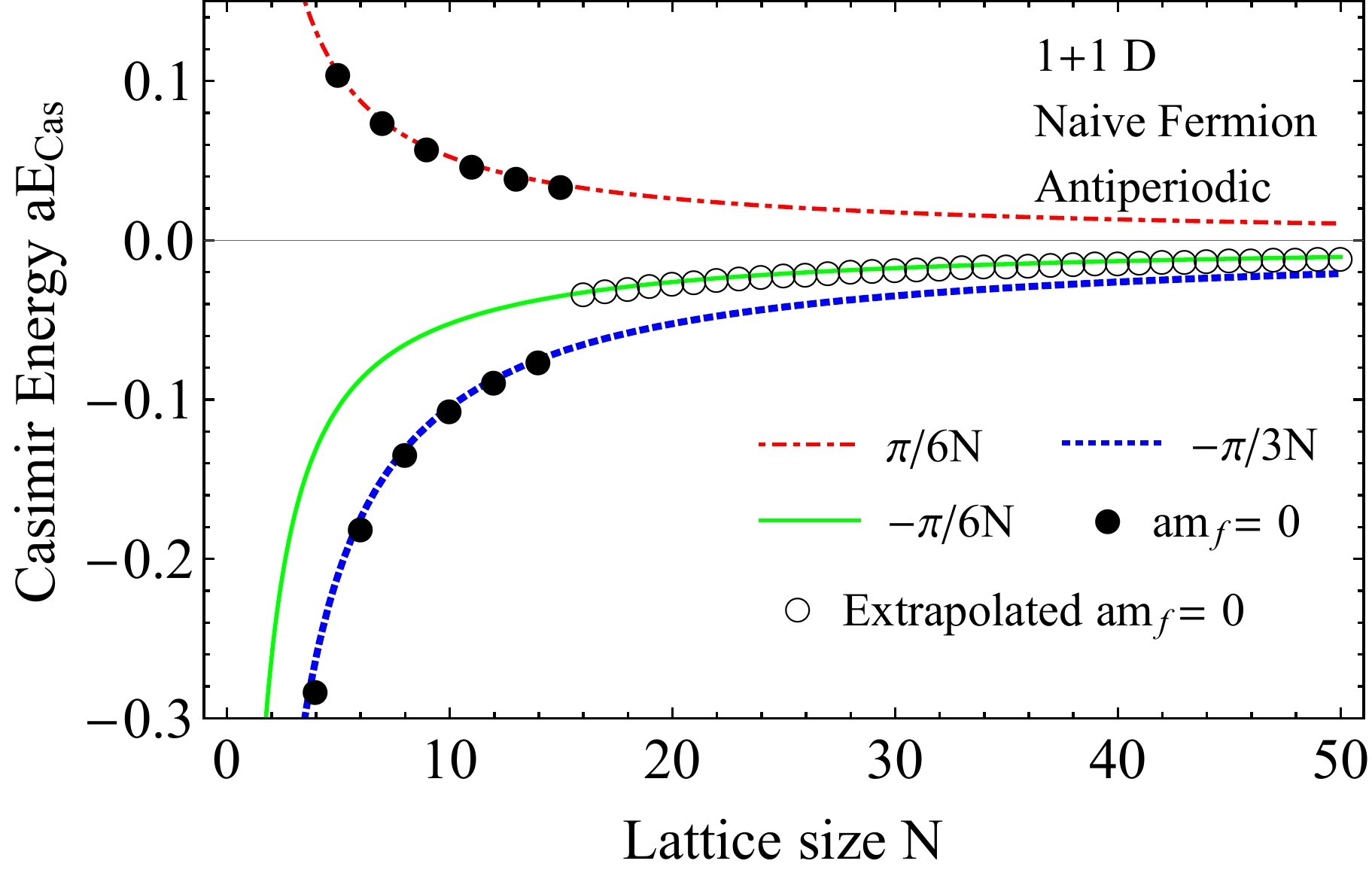} }}%
    \footnotesize{(c)}{{\includegraphics[width=6.25cm]{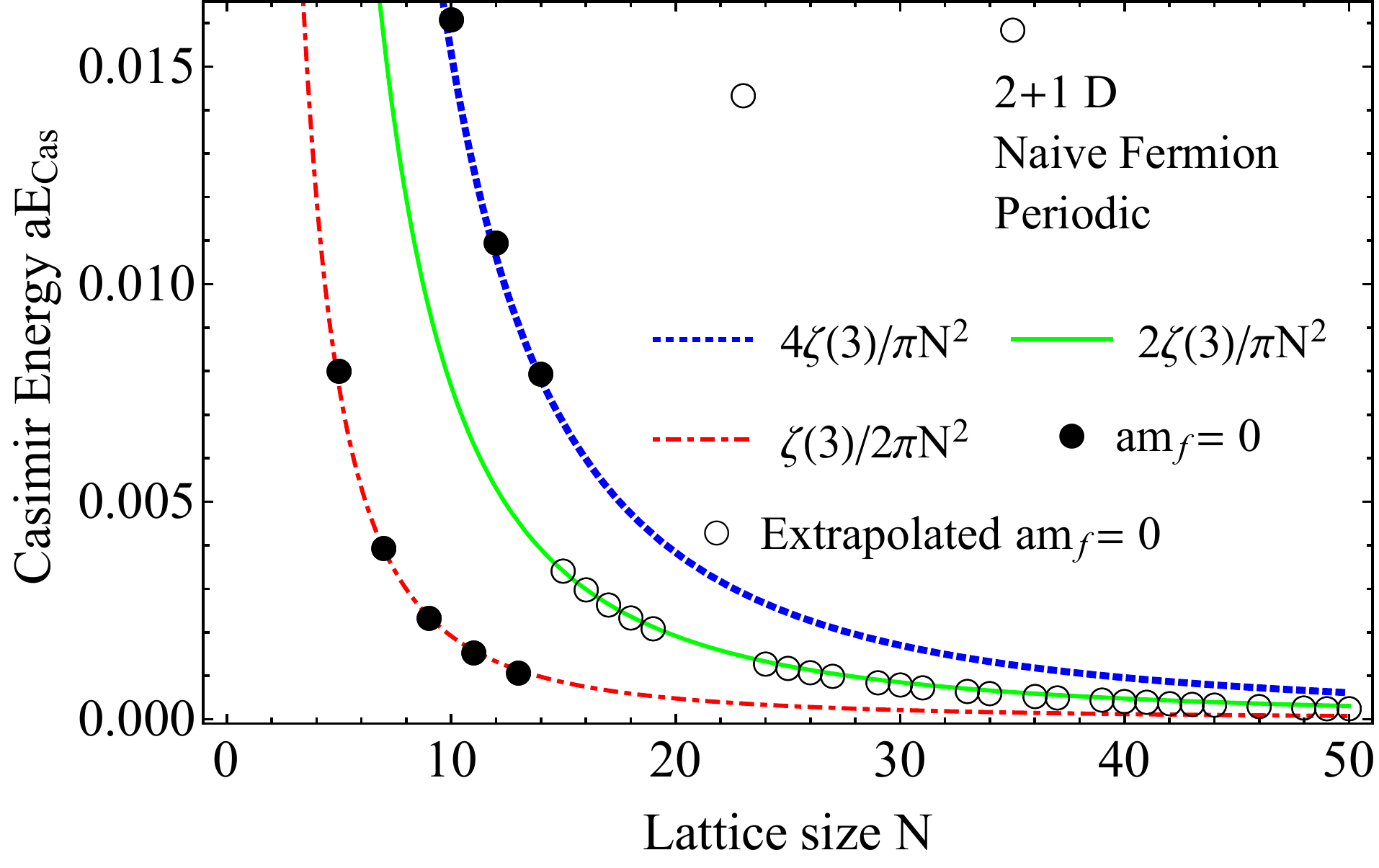}}} %
    \footnotesize{(d)}{{\includegraphics[width=6.25cm]{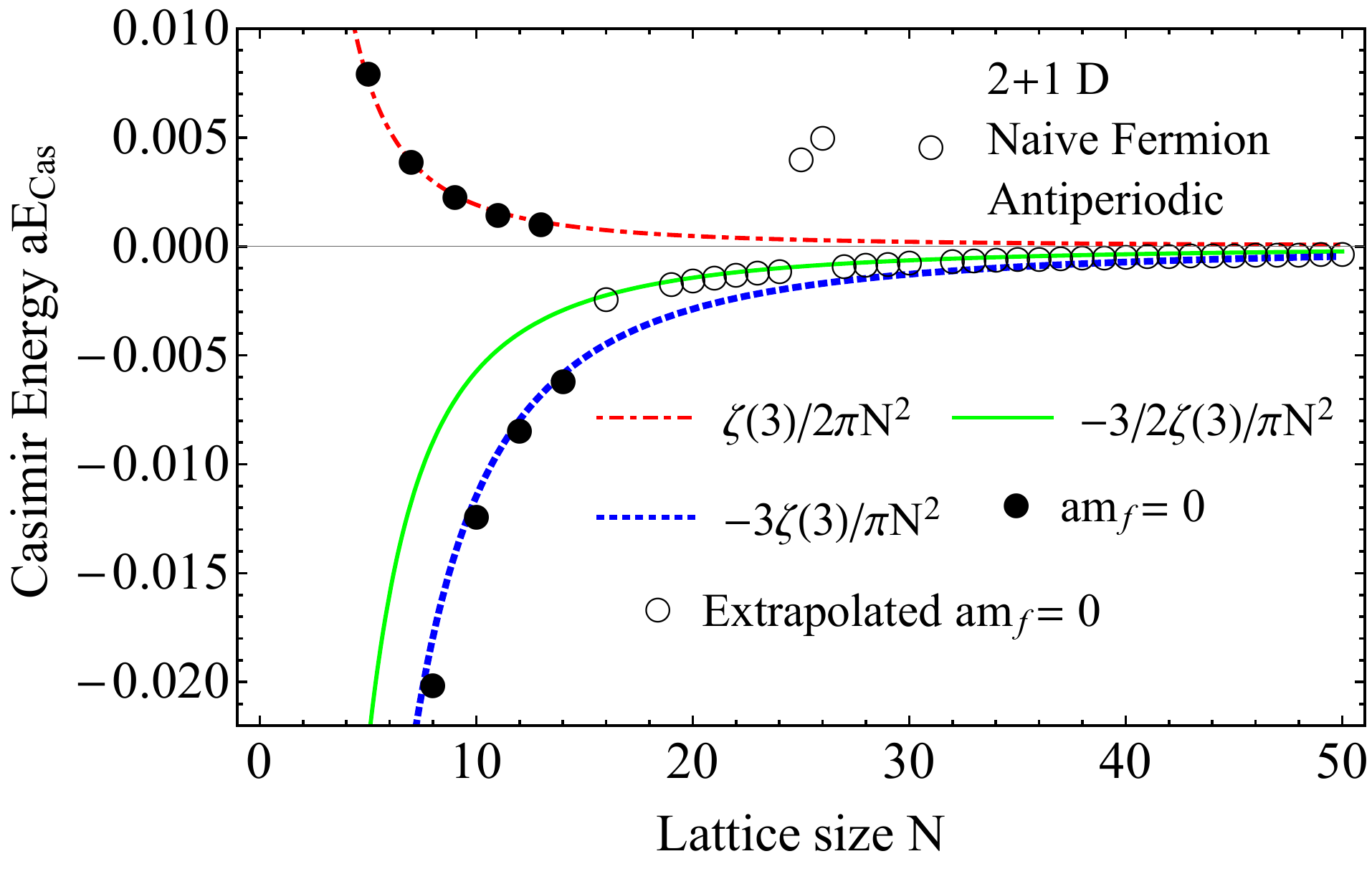}}}
 \caption{\footnotesize{The convergence of Casimir energy for massless naive fermion to the continuum expression after extrapolation is demonstrated by series acceleration methods. [(a),(b)] and [(c),(d)] represent the periodic and antiperiodic boundary conditions for naive fermion in ($1+1$)- and ($2+1$)-dimension respectively.}}%
\label{extrapolation}
\end{figure} 

%Contrary to the claim in Ref. \cite{Ishikawa:2020ezm} that one cannot determine the Casimir effect for Dirac fermions from the continuum limit of the naive fermion formulation, %\textcolor{red}{
Employing the Euler-Maclaurin method as implemented in Wolfram Mathematica, we obtained the results
%that the extrapolated Casimir energy for naive fermion matches the continuum expression for the fermionic Casimir energy precisely in the zero lattice spacing ($a\rightarrow0$) limit
as exhibited in Fig. \ref{extrapolation} (a),(b). The oscillatory Casimir energy data points for $N<15$ are obtained from definition (\ref{def3+1}) for periodic and antiperiodic boundary conditions. The extrapolated results are represented by hollow markers in Fig. \ref{extrapolation} for $N>15$, and signal their precise agreement with the continuum results in (\ref{cont}). The Euler-Maclaurin method of numerical sum extrapolates the data for points greater than the prescribed number to terms to be considered using the \texttt{NSumTerms} attribute of Wolfram Mathematica. In Fig. \ref{extrapolation}, this attribute is set to 15 for the agreement between the extrapolated points and continuum results to be clearly visible. Changing the attribute to higher values, say 20, does not alter the results qualitatively. 
%One can obtain precise results by changing the number of terms to be considered to say, 20 in the Euler-Maclaurin Sum and then extrapolate the result for lattice size $N>20$. 
%give different results for the Casimir energy as calculated in (\ref{naivePcont}, \ref{naiveAPcont}). 
%from these expressions
%We realized that one must not treat the Casimir energy for odd and even $N$ as different cases 
%and expressed by a single, rapidly oscillating function of $N$ as shown by the dashed line in Fig. .
%After our extrapolation of oscillatory expression on lattice described above converges precisely to the same continuum expression at large $N$, as shown in Fig. \ref{extrapolation}.
%Such extrapolation methods accelerate the convergence between the two separate analytic results obtained for odd and even lattice sizes. 
We find that the Casimir energy for naive fermion converge to a single function which is exactly equal to the continuum result. Moreover, these expressions also match the Wilson and overlap fermion results for both the periodic and antiperiodic boundary conditions, as expected from universality. {In contrast to Ref. \cite{Ishikawa:2020ezm},
%}
%The Casimir energy calculations for the naive fermion using periodic and anti-periodic boundary conditions are done with the Abel-Plana formulae in finite range 
%(\ref{int}, \ref{nonint})
%in ($1+1$)-dimensions, and numerically in higher dimensions gave us different continuum expressions for odd and even lattice sizes ($N$) in (\ref{naivep}, \ref{naivea}). 
we therefore argue that the continuum limit of the Casimir energy expression (\ref{naivemitbagexp}) must be treated as a combined oscillatory expression of the odd and even lattice sizes ($N$).}
%Using the above-mentioned series acceleration techniques, we have shown that the oscillating naive Casimir energy converges to a single expression in the continuum limit for sufficiently large $N$ irrespective of the boundary condition and does not violate the universality of lattice fermions in the continuum limit. 
%as it is possible to derive the Dirac fermionic Casimir energy using the naive fermion results, irrespective of the boundary condition, just as we did for Wilson and overlap fermion in this thesis. 
%In Fig. \ref{extrapolation}, one can see that the expression for Casimir energy of naive fermion in (\ref{naivep}, \ref{naivea}) converges to a single expression after extrapolation. 
%This ensures that the universality of lattice fermions is not violated. 

A similar extrapolation works for the $(2+1)$-dimensions as shown in Fig \ref{extrapolation} (c),(d). However,
%from Fig. \ref{extrapolation}(c),(d)
some lattice points are seen to be  off the mark as the numerical computation involves calculating the integral of a  sum approximated by an expression of higher-order derivatives of the function $|\sin(2\pi n/N)|$, over a continuous variable. This function is not differentiable at the lattice points and thus leads to this discrepancy. Nonetheless, it is sufficiently clear that the oscillating series indeed converges to the same continuum expression as $N$ increases.
 
In \cite{Ishikawa:2020icy}, it was shown that the negative mass Wilson fermions behave differently according to phases of the fermion mass and represent the bulk fermions, whereas the overlap fermions represent the surface fermions in a topological insulator. They also exhibit oscillation of Casimir energy on odd and even lattice sizes in certain phases. In all such phases, %where oscillation of Casimir energy is observed for odd and even $N$,
it is observed that these oscillatory results approach different expressions in the continuum limit. We have verified that applying the same principle of treating both odd and even $N$ as part of the same series, and applying the Euler-Maclaurin series extrapolation method for the Casimir energy of the negative mass Wilson fermions also leads to the correct result in the continuum limit. Therefore, it is shown that the apparent universality violations in all known cases so far are cured when the rapid oscillations between odd and even $N$ are treated as a single oscillatory function.

\section{Conclusion}
In this paper, we first studied
%the causes and effects of 
the continuum limit of the Casimir effect for common lattice 
%photons and
Dirac fermions in the light of the MIT Bag model. 
%The fermionic Casimir effect includes modelling a parallel plate slab with the MIT Bag Model and corresponding boundary conditions to calculate the Casimir energy.
In section (\ref{slabbag}), the MIT Bag boundary conditions were realized on a lattice for the first time and used to calculate the Casimir effect for lattice fermions as per the formalism developed in Ref. \cite{Ishikawa:2020ezm}. Following the studies done in Ref. \cite{Ishikawa:2020ezm, Ishikawa:2020icy}, analytic expressions for the Casimir energy of naive (with doubling correction) and Wilson fermions are obtained, along with the numerical results for overlap fermions with MIT Bag boundary conditions.
%with \textcolor{red}{MDW} kernel
%in ($1+1$)-dimensions and higher dimensional spacetime 
These expressions matched the continuum results for Dirac fermions exactly as expected from universality. Results in higher dimensions are also similar to the ($1+1$)-dimensions.
%But, we observe that the Casimir energy oscillation in the naive fermion case, as seen for the periodic and antiperiodic boundary conditions are absent for the MIT Bag boundary conditions.
Moreover, unlike the Casimir energy oscillations in the naive fermion case, seen for both periodic and antiperiodic boundary conditions, MIT Bag boundary conditions do not exhibit any such oscillations, neither do the Dirac fermions in the continuum, nor the other lattice fermions like Wilson and overlap fermions. Interestingly, negative mass Wilson fermions also exhibit such oscillatory behaviour \cite{Ishikawa:2020icy} for a certain range of mass parameter $am_f$, suggesting it to be possibly related to specific structure of the dispersion relations.
%No such oscillations are encountered in Casimir energy calculated for Dirac fermions in continuum or for lattice fermions except the naive fermion. This seemed to suggest an apparent violation of universality for this case. 
%The reason for this non-monotonous behaviour for naive fermion is not the periodic/ antiperiodic boundary conditions, as other lattice fermions agree well with the continuum results with these boundary conditions in the zero-lattice spacing limit. 
%, but in the case of negative mass Wilson fermions, we studied certain special properties like (\ref{nocasimir}) specific to the ($1+1$)- dimensions. 

In Ref. \cite{Ishikawa:2020ezm, Ishikawa:2020icy}, the analytic expressions of Casimir energy for odd and even lattice sizes for naive fermion were obtained and interpreted separately. This led to different results for odd and even lattice sizes, which in turn did not agree with the continuum results, suggesting an apparent universality violation. We noticed that the difference between the Casimir energy for odd and even lattice sizes given in (\ref{diff}), converges rapidly. This led us to propose that the results obtained from two analytic expressions formed a single oscillatory series. In the continuum limit of this single oscillatory series, we obtained the same result as for other lattice fermions agreeing with the continuum results, thus restoring universality. Thus, Casimir energy for naive fermions can also be used to obtain results in the continuum limit, in contrast to the claim in \cite{Ishikawa:2020ezm, Ishikawa:2020icy}.
%Following the studies done in Ref. \cite{Ishikawa:2020ezm, Ishikawa:2020icy}, exact analytic expressions for Casimir energy in continuum, for naive lattice fermions using periodic and antiperiodic boundary conditions were also studied. 
%We have shown that the universality of lattice fermion actions is not violated while calculating the Casimir energy for naive fermion. We proposed that the odd and even lattice sizes are part of the single oscillatory series for these boundary conditions, in contrast to the claim \cite{Ishikawa:2020ezm, Ishikawa:2020icy}, where the continuum limit for the odd and even lattices were taken separately leading to different continuum results for both. 
Employing the Euler-Maclaurin series extrapolation method, better suited for strongly oscillating series like the one obtained, we showed the convergence of the oscillating analytic expression between odd and even lattice sizes to the same expression as continuum. This makes even the naive fermions with periodic/antiperiodic boundary conditions suitable for such calculations.%One can also verify the repulsive Casimir energy between Chern Insulators with oppositely signed Chern numbers \cite{Rodriguez} using this formalism.% This leads us to conclude that the oscillation for odd and even lattice sizes is also dependent on the boundary conditions.
\section{Acknowledgements}
RVG gratefully acknowledges the support of the Department of Atomic Energy, Government of India through Raja Ramanna Fellowship. YVM gratefully acknowledges the support of the INSPIRE fellowship, Department of Science and Technology, Government of India.
\textcolor{red}{
%we realized the MIT Bag boundary conditions for lattice fermions and calculated the corresponding Casimir energy. 
%The expression obtained in the continuum limit for the MIT Bag boundary conditions with a doubling parameter of two in ($1+1$)-dimensions matches the continuum expression exactly.
}
 
\nocite{*}
%\appendix

%\section{Sample Appendix Section}

 \bibliographystyle{elsarticle-num} 
 \bibliography{MAIN}

\begin{thebibliography}{10}
\expandafter\ifx\csname url\endcsname\relax
  \def\url#1{\texttt{#1}}\fi
\expandafter\ifx\csname urlprefix\endcsname\relax\def\urlprefix{URL }\fi
\expandafter\ifx\csname href\endcsname\relax
  \def\href#1#2{#2} \def\path#1{#1}\fi

\bibitem{Ishikawa:2020ezm}
T.~Ishikawa, K.~Nakayama, K.~Suzuki, {Casimir effect for lattice fermions},
  Phys. Lett. B 809 (2020) 135713.
\newblock \href {http://arxiv.org/abs/2005.10758} {\path{arXiv:2005.10758}},
  \href {https://doi.org/10.1016/j.physletb.2020.135713}
  {\path{doi:10.1016/j.physletb.2020.135713}}.

\bibitem{Casimir:1948dh}
H.~B.~G. Casimir, {On the Attraction Between Two Perfectly Conducting Plates},
  Indag. Math. 10 (1948) 261--263.

\bibitem{Bressi}
G.~Bressi, G.~Carugno, R.~Onofrio, G.~Ruoso, Measurement of the casimir force
  between parallel metallic surfaces, Phys. Rev. Lett. 88 (2002) 041804.
\newblock \href {https://doi.org/10.1103/PhysRevLett.88.041804}
  {\path{doi:10.1103/PhysRevLett.88.041804}}.

\bibitem{Lamoreaux}
S.~K. Lamoreaux, Demonstration of the casimir force in the 0.6 to
  $6\ensuremath{\mu}m$ range, Phys. Rev. Lett. 78 (1997) 5--8.
\newblock \href {https://doi.org/10.1103/PhysRevLett.78.5}
  {\path{doi:10.1103/PhysRevLett.78.5}}.

\bibitem{Mohideen}
U.~Mohideen, A.~Roy, Precision measurement of the casimir force from 0.1 to
  $0.9{\mu}m$, Phys. Rev. Lett. 81 (1998) 4549--4552.
\newblock \href {https://doi.org/10.1103/PhysRevLett.81.4549}
  {\path{doi:10.1103/PhysRevLett.81.4549}}.

\bibitem{Gong}
T.~Gong, M.~R. Corrado, A.~R. Mahbub, C.~Shelden, J.~N. Munday, Recent progress
  in engineering the casimir effect – applications to nanophotonics,
  nanomechanics, and chemistry, Nanophotonics 10~(1) (2021) 523--536.
\newblock \href {https://doi.org/10.1515/nanoph-2020-0425}
  {\path{doi:10.1515/nanoph-2020-0425}}.

\bibitem{Johnson:1975zp}
K.~Johnson, {The M.I.T. Bag Model}, Acta Phys. Polon. B 6 (1975) 865.

\bibitem{Milton:2001yy}
K.~A. Milton, {The Casimir effect: Physical manifestations of zero-point
  energy}, World Scientific, 2001.
\newblock \href {https://doi.org/10.1142/4505} {\path{doi:10.1142/4505}}.

\bibitem{ChernodubNonPert}
M.~N. Chernodub, V.~A. Goy, A.~V. Molochkov, {Nonperturbative Casimir Effects
  in Field Theories: aspects of confinement, dynamical mass generation and
  chiral symmetry breaking}, PoS Confinement2018 (2019) 006.
\newblock \href {http://arxiv.org/abs/1901.04754} {\path{arXiv:1901.04754}},
  \href {https://doi.org/10.22323/1.336.0006} {\path{doi:10.22323/1.336.0006}}.

\bibitem{Morris}
M.~S. Morris, K.~S. Thorne, U.~Yurtsever, Wormholes, time machines, and the
  weak energy condition, Phys. Rev. Lett. 61 (1988) 1446--1449.
\newblock \href {https://doi.org/10.1103/PhysRevLett.61.1446}
  {\path{doi:10.1103/PhysRevLett.61.1446}}.

\bibitem{MAHAJAN20066}
G.~Mahajan, S.~Sarkar, T.~Padmanabhan, Casimir effect confronts cosmological
  constant, Physics Letters B 641~(1) (2006) 6--10.
\newblock \href
  {https://doi.org/https://doi.org/10.1016/j.physletb.2006.08.026}
  {\path{doi:https://doi.org/10.1016/j.physletb.2006.08.026}}.

\bibitem{Araki:2013qva}
Y.~Araki, T.~Kimura, A.~Sekine, K.~Nomura, T.~Z. Nakano, {Phase structure of
  topological insulators by lattice strong-coupling expansion}, PoS LATTICE2013
  (2014) 050.
\newblock \href {http://arxiv.org/abs/1311.3973} {\path{arXiv:1311.3973}},
  \href {https://doi.org/10.22323/1.187.0050} {\path{doi:10.22323/1.187.0050}}.

\bibitem{Rodriguez}
P.~Rodriguez-Lopez, A.~G. Grushin, Repulsive casimir effect with chern
  insulators, Phys. Rev. Lett. 112 (2014) 056804.
\newblock \href {https://doi.org/10.1103/PhysRevLett.112.056804}
  {\path{doi:10.1103/PhysRevLett.112.056804}}.

\bibitem{Ishikawa:2020icy}
T.~Ishikawa, K.~Nakayama, K.~Suzuki, {Lattice-fermionic Casimir effect and
  topological insulators}, Phys. Rev. Res. 3~(2) (2021) 023201.
\newblock \href {http://arxiv.org/abs/2012.11398} {\path{arXiv:2012.11398}},
  \href {https://doi.org/10.1103/PhysRevResearch.3.023201}
  {\path{doi:10.1103/PhysRevResearch.3.023201}}.

\bibitem{Susskind}
L.~Susskind, {Lattice Fermions}, Phys. Rev. D 16 (1977) 3031--3039.
\newblock \href {https://doi.org/10.1103/PhysRevD.16.3031}
  {\path{doi:10.1103/PhysRevD.16.3031}}.

\bibitem{Mamaev:1980jn}
S.~G. Mamaev, N.~N. Trunov, {Vacuum expectation values of the energy-momentum
  tensor of quantized fields on manifolds with different topologies and
  geometries III}, Sov. Phys. J. 23 (1980) 551--554.
\newblock \href {https://doi.org/10.1007/BF00891938}
  {\path{doi:10.1007/BF00891938}}.

\bibitem{BelussiSaharian}
S.~Bellucci, A.~A. Saharian, Fermionic casimir effect for parallel plates in
  the presence of compact dimensions with applications to nanotubes, Phys. Rev.
  D 80 (2009) 105003.
\newblock \href {https://doi.org/10.1103/PhysRevD.80.105003}
  {\path{doi:10.1103/PhysRevD.80.105003}}.

\bibitem{Paola}
R.~D.~M. Paola, R.~B. Rodrigues, N.~F. Svaiter, {Casimir} {Energy} {of}
  {Massless} {Fermions} {in} {the} {Slab}-{Bag}, Modern Physics Letters A
  14~(34) (1999) 2353--2361.
\newblock \href {https://doi.org/10.1142/s0217732399002431}
  {\path{doi:10.1142/s0217732399002431}}.

\bibitem{ELIZALDE}
E.~Elizalde, F.~C. Santos, A.~C. Tort, {The} {Casimir} {energy} {of} a
  {Massive} {Fermionic} {field} {confined} {in} a ($d + 1$)-{dimensionsal}
  {Slab}-{Bag}, International Journal of Modern Physics A 18~(10) (2003)
  1761--1772.
\newblock \href {https://doi.org/10.1142/s0217751x03014186}
  {\path{doi:10.1142/s0217751x03014186}}.

\bibitem{Cruz}
M.~B. Cruz, E.~R.~B. de~Mello, A.~Y. Petrov, Fermionic casimir effect in a
  field theory model with lorentz symmetry violation, Phys. Rev. D 99 (2019)
  085012.
\newblock \href {https://doi.org/10.1103/PhysRevD.99.085012}
  {\path{doi:10.1103/PhysRevD.99.085012}}.

\bibitem{Gattringer:2010zz}
C.~Gattringer, C.~B. Lang, {Quantum chromodynamics on the lattice}, Vol. 788,
  Springer, Berlin, 2010.
\newblock \href {https://doi.org/10.1007/978-3-642-01850-3}
  {\path{doi:10.1007/978-3-642-01850-3}}.

\bibitem{Saharian:2006iv}
A.~A. Saharian, {Generalized Abel-Plana formula as a renormalization tool in
  quantum field theory with boundaries}, PoS IC2006 (2006) 019.
\newblock \href {http://arxiv.org/abs/hep-th/0609093}
  {\path{arXiv:hep-th/0609093}}, \href {https://doi.org/10.22323/1.031.0019}
  {\path{doi:10.22323/1.031.0019}}.

\bibitem{Saharian:2007ph}
A.~A. Saharian, The generalized abel-plana formula with applications to bessel
  functions and casimir effect, arXiv: High Energy Physics - Theory (2007).

\end{thebibliography}





% \end{thebibliography}
\end{document}